\documentclass[12pt]{article}
\usepackage{setspace}
\usepackage{multirow}
\usepackage{amsfonts}
\usepackage{amssymb}
\usepackage{graphicx}
\usepackage{amsmath}
\usepackage{harvard}
\usepackage{environ}
\usepackage{pdfpages}
\usepackage{bbm}
\usepackage{caption}
\usepackage{multirow}

\newtheorem{theorem}{Theorem}

\newtheorem{corollary}{Corollary}

\newtheorem{proposition}{Proposition}

\newenvironment{proof}[1][Proof]{\textbf{#1.} }{\ \rule{0.5em}{0.5em}}

\renewcommand{\cite}{\citeasnoun}
\newif\ifhideproofs
\ifhideproofs
\NewEnviron{hide}{}

\fi
\begin{document}

\author{Dirk Bergemann\thanks{%
Yale University, dirk.bergemann@yale.edu.} \and Tibor Heumann\thanks{%
Instituto de Econom\'{\i}a, Pontificia Universidad Cat\'{o}lica de Chile,
tibor.heumann@uc.cl.} \and Stephen Morris\thanks{%
Massachusetts Institute of Technology, semorris@mit.edu}}
\title{Information Design and Mechanism Design: \linebreak An Integrated
Framework\thanks{%
We acknowledge financial support from NSF grants SES-2001208 and
SES-2049744, ONR-MURI N00014-24-1-2742 and ANID Fondecyt 1241302. We thank
the audience at the World Congress of the Econometric Society \ in Seoul and
the discussant Anton Kolotilin. We benefited from discussions with Ben
Brooks, Roberto Carrao, Ian Jewitt, Benny Moldovanu, Georg N\"{o}ldeke,
Philipp Strack, Daniel Quigley, Alexander Wolitzky and Qinyu Wu. We thank
Marek Bojko and David Wambach for excellent research assistance.} }
\date{\today }
\maketitle

\begin{abstract}
We develop an integrated framework for information design and mechanism
design in screening environments with quasilinear utility. Using the tools
of majorization theory and quantile functions, we show that both information
design and mechanism design problems reduce to maximizing linear functionals
subject to majorization constraints. For mechanism design, the designer
chooses allocations weakly majorized by the exogenous inventory. For
information design, the designer chooses information structures that are
majorized by the prior distribution. When the designer can choose both the
mechanism and the information structure simultaneously, then the joint
optimization problem becomes bilinear with two majorization constraints. We
show that pooling of values and associated allocations is always optimal in
this case. Our approach unifies classic results in auction theory and
screening, extends them to information design settings, and provides new
insights into the welfare effects of jointly optimizing allocation and
information.

\noindent \textsc{Jel Classification: }D44, D47, D82, D83.

\end{abstract}

\newpage

\section{Introduction}

We offer an integrated perspective on mechanism design and information
design, emphasizing the deep symmetry between these two classes of problems.
We use the geometric language of majorization to provide a unified treatment
that reveals structural parallels of information design and mechanism design.

We start with the classic screening problem of \cite{muro78}, where a
monopolist sells goods of varying quality to buyers with heterogeneous
values. Following \cite{lomu22}, we assume the seller has an exogenous
inventory of goods rather than producing them at convex cost. Instead of
working directly with distributions of the values of the buyer and the
qualities of the good, we work throughout with their inverses---that is,
quantile functions. Thus, we describe the problem in terms of a quantile
value function $V(t)$, giving the value $V\,$of the buyer at the $t$-th
quantile, and a quantile allocation function $Q(t)$, giving the quality $Q$\
of the good at the $t$-th quantile.

The core contribution of this paper is to show how three fundamental
problems in pricing and information provision can be understood as
variations of a single mathematical structure. We provide an envelope
formula expressing the seller's revenue as a function of quantile value and
allocation functions. This bilinear representation allows us to analyze
three distinct optimization problems:

First, we characterize the optimal screening mechanism for an exogenous
information structure. The seller chooses a quantile allocation function
that is weakly majorized by the exogenous inventory to maximize revenue.
This turns the objective function of the seller into a linear functional of
the allocation. This approach follows \cite{klms21} who emphasized the role
of majorization constraints in economic optimization problem (see also \cite%
{klms26}, their survey article in this volume). The solution involves
concavifying an appropriately defined pointwise revenue function, naturally
leading to pooling (or "ironing") in regions where monotonicity constraints
bind.

Second, we consider the information design problem: choosing the optimal
information structure given an exogenous mechanism that allocates goods
efficiently. Here, the designer maximizes a linear functional subject to a
(strict) majorization constraint---the set of feasible information
structures consists of all mean-preserving contractions of the prior
distribution of values. Remarkably, this problem mirrors the mechanism
design exercise. The solution again involves a concavification procedure,
but now applied to a different function and with a different majorization
constraint.

Third, we examine the problem where the designer can choose both the
information structure and the mechanism jointly. The optimization is now
bilinear in the quantile value and quantile allocation functions, subject to
two majorization constraints. We show that pooling information is always
optimal when both dimensions are jointly chosen. By pooling information, we
refer to any information structure that provides less than complete
information to the buyer about their value; however, coarse information
translates directly into the optimality of a mechanism that offers few
distinct items. This encompasses any deterministic or stochastic information
structure that sends the same signal for at least two distinct values. This
result contrasts sharply with the separate optimization problems, where full
disclosure or full separation can be optimal depending on the primitives.

Throughout the paper, we relate the screening problem to the classic single
good auction setting with a finite number of bidders. \cite{klms21} noted
that when one restricts attention to symmetric bidder distributions and
symmetric mechanisms, the optimal auction problem is equivalent to the
screening problem via a generalization of Border's theorem (\cite{bord91}).
Specifically, in an auction with $N$ bidders, the relevant quantile
allocation function is $Q(t)=t^{N-1}$. All three of our optimization
exercises extend naturally to this auction interpretation.

In \cite{bhms22}, we studied the information design problem in the auction
case, showing that the optimal information structure depends only on the
number of bidders and involves upper censoring. In \cite{behm26}, we
analyzed the screening problem where the seller can choose both the
mechanism and information structure. The auction version of this problem was
earlier analyzed by \cite{bepe07}; the symmetric case of the auction problem
can thus be understood as a special case of our screening framework.

Our approach differs from the standard Bayesian persuasion framework of \cite%
{kage11}, where the sender's payoff is a function of the receiver's
posterior expected value (state). In the Bayesian persuasion setting, there
exists a well-known characterization of the distribution of posterior
beliefs that can be generated by any information structure: namely, the ones
that are a mean-preserving contractions of the distribution of values. The
characterization of the set of feasible distributions of expected values
remains valid in our information design problem; however, it is not useful
because our problem is linear in expected values but nonlinear in the
corresponding probabilities. To get tractability, we work in the quantile
space. Thus, while the objective function is the same as in the classic
Bayesian persuasion framework, the distributions of expected values are
replaced with the quantile distributions. Importantly, the set of feasible
quantile distributions is given by the mean-preserving spreads of the prior
quantile value function. Hence, we invert the direction of the constraint by
moving from value distributions to quantile distributions. Methodologically,
we follow \cite{klms21} in their work on majorization constraints and
therefore also use the corresponding terminology. Interestingly, this places
our information design closer to the \cite{myer81} ironing technique, which
also employed quantile functions. Recent work by \cite{jequ23} shows that
our problem is equivalent to Bayesian persuasion with \cite{yaar87}
preferences---linear in outcomes but non-linear in probabilities.

While our primary focus is revenue maximization, we also examine alternative
objectives. We show that the consumer surplus maximization problem exhibits
a natural duality with revenue maximization, allowing us to apply the same
geometric techniques. We characterize the complete welfare frontier---the
set of all feasible revenue and consumer surplus pairs under all possible
information structures. For the auction case with an efficient allocation
rule, we show that the frontier can be traced out by upper and lower
censoring information structures, with the optimal structure switching
discontinuously at the point where total surplus is maximized. \ These
results illustrate the robust predictions approach described in \cite%
{bemo13a}, (2016) \nocite{bemo16}: we don't think that there is a literal
information designer who is maximizing a weighted sum of revenue and
consumer surplus. \ Rather, the information designer is metaphorical and by
examining information design for all possible weights, we identify which
combinations of revenue and consumer surplus are possible in equilibrium
under some information structure. \ 

We also examine comparative statics in the number of bidders in the auction
problem. We show that the optimal information structure in
revenue-maximizing auctions involves pooling a decreasing fraction of the
value distribution as $N$ grows. However, a striking regularity emerges: the
expected number of bidders whose values are pooled at the top remains
approximately constant at about two bidders, regardless of $N$. This
suggests that optimal information design sustains competition among roughly
two bidders at the margin, even in large auctions.

Our framework connects to several strands of literature beyond those already
mentioned. The problem of jointly optimizing pricing and information relates
to recent work on data markets and digital advertising, see \cite{bhms22}, 
\cite{bebo24}, \cite{bebw25}. The revenue guarantee problem---finding
mechanisms that perform well across all information structures---connects to
the informationally robust mechanism design literature surveyed by \cite%
{brdu25}. Finally, our use of quantile methods links to the virtual value
approach in auction theory by \cite{myer81} and \cite{bukl96}.

Our analysis focuses exclusively on mechanism design with transfers. \cite%
{klms21} provide a connection between information design and delegation
(mechanism design without transfers) in a linear-quadratic environment using
majorization techniques; \cite{koza25} provide a more general connection
between these two classes of problems. \cite{lou23} analyzes the joint
design of delegation and information design. Similarly, \cite{ivan10}
considers the joint choice of information design and information
transmission in the context of a game of strategic communication (cheap
talk). He shows that information control increases the benefit for the
sender relative to a delegation solution in the setting of \cite{crso82}.

The remainder of the paper proceeds as follows. Section \ref{sec:model}
introduces the model and establishes notation for quantile functions,
information structures, and mechanisms. Section \ref{sec:env} derives the
envelope formula for revenue and consumer surplus in quantile space.
Sections \ref{sec:mec}, \ref{sec:infode} and \ref{sec:combine} analyze
mechanism design, information design, and joint design optimization of
allocation and information, respectively. Section \ref{sec:cons} examines
consumer surplus maximization and weighted welfare objectives. Section \ref%
{sec:rob} discusses robust predictions, characterizes the welfare frontier,
and provides comparative statics in the number of bidders.

\section{Setting\label{sec:model}}

\subsection{Quantile Representation of the Screening Problem}

A seller supplies goods of varying quality $q\in \mathbb{R}_{+}$ to a
continuum of buyers with mass $1$ indexed by $t\in \lbrack 0,1]$. Each buyer
has unit demand and a willingness-to-pay, or value, ${v}\in \mathbb{R}_{+}$
for quality $q$. The utility net of the payment $p\in \mathbb{R}_{+}$ is: 
\begin{equation}
{v\cdot }q-p.  \label{uu}
\end{equation}%
The functional form - which follows \cite{muro78} - is important because it
ensures that buyers only care about the expected quality of the good they
are supplied. \ The buyers' values ${v}\in \lbrack \underline{v},\overline{v}%
]$ are distributed according to $F_{v}\in \Delta \left( \lbrack \underline{v}%
,\overline{v}]\right) $ where $0\leq \underline{v}<\overline{v}\leq \infty $%
. Moreover, we can interpret $q$ as the quality, but also a quantity or
probability that a good is provided. In this second interpretation of $q$,
we can link the problem to the auction problem where a single unit is
provided among many symmetric bidders.

We assume that there is an exogenous inventory of the good available to the
seller. \ Specifically, the seller has a mass $1$ of goods with qualities $%
q\in \lbrack \underline{q},\overline{q}]$ distributed according to 
\begin{equation}
F_{q}\in \Delta \left( \lbrack \underline{q},\overline{q}]\right) \text{,}
\label{eq:q}
\end{equation}%
where $0\leq \underline{q}<\overline{q}\leq \infty $. Thus, the seller can
offer a fixed and exogenously given distribution $F_{q}$\ of qualities.

It is without loss of generality to assume that the inventory has mass 1 as
a smaller inventory can be augmented with goods of quality $q=0$. For
example, if the inventory consists of a mass 1/2 of homogeneous goods of
quality $q=1$ (so only 1/2 of the buyers can be served), this corresponds to
the distribution: 
\begin{equation*}
F_{q}(q)=%
\begin{cases}
\frac{1}{2}, & \text{if }q\in \lbrack 0,1); \\ 
1, & \text{if\ }q\geq 1.%
\end{cases}%
\end{equation*}%
Hence, the mass of 1/2 at $q=0$ captures the fact that 1/2 of the buyers
cannot be served.


We denote the inverse of the each distribution, the corresponding quantile
function, as follows: 
\begin{equation}
{V}(t)\triangleq F_{v}^{-1}(t)\quad \text{and}\quad {Q}(t)\triangleq
F_{q}^{-1}(t).  \label{eq:inv}
\end{equation}%
We will refer to $V$ as a quantile value function, which captures the
exogenous distribution of values among buyers. \ We will refer to $Q$ as a
quantile allocation function and it captures the exogenous inventory of the
seller. Whenever one of the distributions has gaps in its support, we adopt
the convention that $V,Q$ are right continuous.

This is the classic screening environment of \cite{muro78}, with the proviso
that, like \cite{lomu22}, we take the inventory of goods sold to be
exogenous. \ This is convenient for our survey. \ The complete analysis of 
\cite{muro78} can then be restored if we endogenize the inventory, i.e.,
allow the seller to produce goods of different distributions of qualities at
some convex cost. \ 

We present our analysis in the quantile space $t\in \left[ 0,1\right] $
rather than the value and quality space $v,q\in \mathbb{R}_{+}$. Mapping to
quantile space is convenient and is done implicitly or explicitly in many
contexts, see\ \cite{myer81}, \cite{hart17}, \cite{bhms22} and \cite{bukl96}.

\subsection{Information Structures}

While $V$ captures the distribution of ex post values in the buyer
population, buyers may be imperfectly informed about their values. \ 

An information structure is given by a mapping $I:[\underline{v},\overline{v}%
]\rightarrow \Delta \left( S\right) $, with the interpretation that $I\left(
v\right) $ is a distribution of signals in some signal space $S$ observed by
buyers with a latent value $v$. A buyer's expected value conditional on the
signal realization $s$ is denoted by: 
\begin{equation}
w\left( s\right) \triangleq \mathbb{E}[{v}\mid s].  \label{w}
\end{equation}%
Importantly, a buyer's expectation of their value will be a sufficient
statistic for their preferences. \ This is an assumption that is widely
adopted in the Bayesian persuasion literature (see \cite{kolo18}).

Frequently, we can omit the dependence of the expected value $w\left(
s\right) $ on the signal $s$ and simply write $w$ for a generic expected
value. We denote by $F_w$ the distribution of expected valuations generated
by an information structure, and denote by $W(t)$ the associated quantile
expected value function: 
\begin{equation*}
W(t)\triangleq F_w^{-1}(t),
\end{equation*}%
which, as before, we adopt the convention that it is right continuous. %
%
%
%
%
%
%
%
%
%
%
%
%
%
%
%
%
%
%
%
We will refer to $W$ as a quantile value function as ${W}(t)$ is the
expected value of the $t$-th quantile buyer in the population.

\subsection{Selling Mechanisms}

A selling mechanism is described by a menu of qualities and payments. The
seller has the ability to shape the distribution of expected qualities sold
by offering lotteries on goods of different qualities. If buyers are offered
a distribution of qualities, or a stochastic bundle $B$ in form of a
lottery, with 
\begin{equation*}
B\in \Delta \left( \lbrack \underline{q},\overline{q}]\right) ,
\end{equation*}%
then the expected quality $x$\ offered is: 
\begin{equation*}
x\triangleq \int_{\underline{q}}^{\overline{\bar{q}}}qdB(q).
\end{equation*}%
If $\int dB(q)<1$, then the probability of receiving an object is less than $%
1$. This is of course equivalent to receiving a zero quality good with
probability $1-\int dB(q)$. The expected utility of a bundle $B$\ net of the
payment $p\in \mathbb{R}_{+}$ is: 
\begin{equation}
{w}\int_{\underline{q}}^{\bar{q}}qdB(q)-p.  \label{u}
\end{equation}%
We now describe the set of possible mechanisms.

A menu indexed by $t\in \lbrack 0,1]$ consists of bundles $B\left( \left.
\cdot \right\vert t\right) $ at prices $p\left( t\right) $:%
\begin{equation}
M\triangleq \left\{ (B\left( \left. \cdot \right\vert t\right)
,p(t))\right\} _{t\in \lbrack 0,1]}.  \label{m}
\end{equation}%
Without loss of generality, we assume that the lotteries with a higher index 
$t$ correspond to higher expected qualities, and denote the quantile
distribution of expected qualities by: 
\begin{equation*}
X\left( t\right) \triangleq \int_{\underline{q}}^{\bar{q}}qdB\left( \left.
q\right\vert t\right) .
\end{equation*}%
Without loss of generality, we assume that-- in addition to being
nondecreasing-- it is right continuous. We say that a menu induces a
quantile distribution of qualities $X(t)$; we formalize later the fact that
this is indeed a quantile distribution. The menu must satisfy the
feasibility constraint. That is, the seller can only bundle and discard
goods, but cannot produce more goods. For now, we omit this constraint and
make it explicit later.

\subsection{The Auction Model}

We will also discuss the translation of our results into a symmetric auction
setting. \ There is a single object to be allocated. \ There are $N$ bidders
who have values drawn independently according to a common distribution $H$.
Now, the probability that a bidder $n$\ with quantile $t$ in the value
distribution $H\ $has the highest value among all bidders is simply ${Q}%
(t)=t^{N-1}$, the probability that $N-1$ bidders have a quantile lower than $%
t$. Note that this expression does not depend on the common distribution $H$%
\ of values. It will turn out that the optimal symmetric auction solution is
essentially identical to that of the screening problem when $Q$ represents
the exogenous inventory of different qualities. Thus, solutions to the
auction model will be identical to solutions for a corresponding screening
model. \ 

Throughout the paper, we will illustrate our results for the case of
symmetric power distributions of values and qualities: 
\begin{equation}
F_{v}\left( v\right) =v^{\frac{1}{4}}\text{,\ \ and }F_{q}\left( q\right)
=q^{\frac{1}{4}}.  \label{eq:pow}
\end{equation}%
The associated quantile functions are given by 
\begin{equation*}
V\left( t\right) =Q\left( t\right) =t^{4}.
\end{equation*}
\ In the case of the auction model, this corresponds to the case of five
bidders i.e., $N=5$ competing for a single unit of an object.

\section{Revenue and Consumer Surplus: \newline
The Envelope Theorem in Quantile Space\label{sec:env}}

We begin our analysis by finding the consumer surplus and profits generated
by an arbitrary information structure and mechanism. For now, we abstract
away from the signals or the lotteries in a menu, and instead write the
consumer surplus and profits in terms of the induced quantile (expected)
value function $W(t)$ and the induced quantile (expected) allocation
function $X(t).$

A buyer with expected value $W(t)$ will choose a quantile $t^{\prime }$ and
associated item $X\left( t^{\prime }\right) $ in the mechanism to solve the
following: 
\begin{equation*}
u(t)\triangleq \max_{t^{\prime }}\left\{ W(t)X(t^{\prime })-p(t^{\prime
})\right\} .
\end{equation*}%
By the revelation principle, it is without loss of generality to assume that
the mechanism is incentive compatible, so that $t^{\prime }=t$ is always the
optimal choice for the consumer with the value $W(t)$. Following the
Envelope theorem we find that $u(t)$ is determined by the allocation $X(t)$
as follows: 
\begin{equation*}
du\left( t\right) =X\left( t\right) dW\left( t\right) ,
\end{equation*}%
and in integral form 
\begin{equation*}
u(t)=\int_{0}^{t}X(s)dW(s)+u(0).
\end{equation*}%
As is standard, the lowest type is always left with zero consumer surplus,
so we have $u(0)=0$. We thus have that the expected consumer surplus is
given by: 
\begin{equation}
\mathbb{E}[u(t)]=\int_{0}^{1}\left[ \int_{0}^{t}X(s)dW(s)\right]
dt=\int_{0}^{1}(1-t)X(t)dW(t).  \label{eq:con}
\end{equation}%
The first equality is obtained by noting that the quantiles are always
uniformly distributed in $[0,1]$; the second equality is obtained by
integrating by parts.

The payment by a $t$-quantile buyer is given by the total surplus generated
by this buyer less their indirect utility, so its given by: 
\begin{equation}
p(t)=W(t)X(t)-\int_{0}^{t}X(s)dW(s).  \label{Eq:obj1}
\end{equation}%
The expected revenue of the seller is thus given by: 
\begin{equation}
\mathbb{E}[p(t)]=\int_{0}^{1}\bigg(W(t)X(t)-\int_{0}^{t}X(s)dW(s)\bigg)%
dt=\int_{0}^{1}(1-t)W(t)dX(t)+X(0)W(0).  \label{eq:rev}
\end{equation}%
as before, the first equality is obtained by noting that the quantiles are
always uniformly distributed in $[0,1]$; the second equality is obtained
integrating by parts twice.

We thus denote the expected consumer surplus and expected revenue by: 
\begin{eqnarray}
U(W,X) &\triangleq &\mathbb{E}[u(t)],\ \ \ \ \ \   \label{eq:ur} \\
\ \ \ \ \ R(W,X) &\triangleq &\mathbb{E}[p(t)].  \label{eq:re}
\end{eqnarray}%
where the notation now emphasizes that revenue is solely dependent on the
quantile value function (the information structure) and the quantile
allocation function (the mechanism).

There are two important aspects of the description that we have provided in
terms of quantile functions. First, revenue and consumer surplus are
bilinear in the quantile functions. Second, the expression for revenue and
consumer surplus show a remarkable symmetry in (\ref{eq:con}) and (\ref%
{eq:ur}). The integrator $W\left( t\right) $ in the revenue is the integrand
in the consumer surplus. Conversely, the integrator $X\left( t\right) $ in
the consumer surplus is the integrand in the revenue. Moreover, if 
\begin{equation}
X\left( 0\right) =0\text{ or }W\left( 0\right) =0,  \label{eq:bou}
\end{equation}%
and this boundary condition typically holds as either the lowest type has
zero value or receives zero allocation, then we have a complete payoff
symmetry in that 
\begin{equation}
R\left( W,X\right) =U\left( X,W\right) .  \label{eq:psy}
\end{equation}%
Thus, the revenue with quantile value function $W\left( t\right) $ and
allocation quantile function $X\left( t\right) $ is the same as the consumer
surplus if $X\left( t\right) $ is the quantile value function and $W\left(
t\right) $ is the quantile allocation function. We shall return to the
symmetry in particular in Section \ref{sec:cons} when we consider solutions
that focus on maximizing consumer surplus. Third, by expressing the revenue
(or consumer surplus) directly in terms of the quantile allocation function
and the quantile value function, the feasibility of the allocation rule
becomes uncoupled from the feasibility of the information structure. In
other words, the feasibility of either allocation or information policy
become uncoupled. We will explore the usefulness of this uncoupling
immediately in the analysis of the optimal mechanism.





\section{Mechanism Design\label{sec:mec}}

We first identify the revenue-maximization mechanism, fixing an exogenous
information structure $W\left( t\right) $. We characterized revenue for
arbitrary information structures in the previous section, but our leading
interpretation in this section is that buyers know their values and so $%
W(t)=V(t)$.

We begin the analysis by characterizing the set of feasible distributions of
qualities that the seller can generate across all mechanisms. We show that
an allocation is feasible if and only if the induced quantile allocation
function $X$ is weakly majorized by the exogenous inventory $Q$. The
seller's revenue can then be expressed as a linear functional of $X$, and
the optimal mechanism is characterized by the concavification of the
pointwise revenue function, which we will define in due course. We then
interpret the result in the context of auctions, provide an example and
examine some of the classic results that establish when the optimal
mechanism does not exhibit pooling. We conclude this section by providing a
general discussion about the results we present and the corresponding
literature.

\subsection{Feasibility of Allocation}

Following \cite{klms21} (see Proposition 4), an allocation rule--which is
mapping from quantiles to qualities-- is feasible if and only if the
quantile allocation rule $X$ satisfies 
\begin{equation}
\int_{x}^{1}X({t})dt\leq \int_{x}^{1}Q({t})dt\text{, }\forall {x}\in \lbrack
0,1],  \label{eq:wm}
\end{equation}%
where we do not require that the inequality becomes an equality at $x=0$. In
this case, we say $Q$ \emph{weakly majorizes} $X$ and we write $X\prec _{w}Q$
(the weak majorization order is equivalent to the increasing convex order).
As discussed above, the seller has the ability to pool goods of different
quality, which corresponds to choosing distributions $X$ that are a mean
preserving contraction of the distribution $Q$. \ However, the seller has
the ability not to sell some products, which means that the qualities of
these products are set to $0$. This is equivalent to allowing downward
shifts in the distribution of qualities in the sense of first-order
stochastic dominance. And, in fact, for any $X\prec _{w}Q$, there exists a
distribution of qualities $\widehat{X}$ such that $\widehat{X}$ is a
mean-preserving contraction of $Q$ and $\widehat{X}$ first-order
stochastically dominates $X$ (see \cite{shsh07}, Theorem 4.A.6.). Thus this
part of the seller's problem corresponds to first pooling and then reducing
qualities.

Here we emphasize a distinct advantage of our formulation in terms of
quantile functions. The feasibility constraint stated in (\ref{eq:wm}) --
when expressed as a quantile function-- does not depend on the information
structure. If instead we were to consider an incentive compatible menu where
the allocation rule is expressed as a function of the value, then the
distribution of goods allocated by the mechanism will obviously depend on
the distribution of (expected) values. Hence, the feasibility constraint
becomes \emph{coupled} with the information structure (and with the
underlying distribution of values).

\paragraph{Feasibility in Auctions}

When the seller chooses the optimal selling mechanism of an indivisible
object with $N$ bidders, the allocation rule is a function $%
q_{n}(v_{n},v_{-n})$ that determines the probability that agent $n$ wins the
object given the report of all bidders. The winning probability is $%
q_{n}(v_{n})=\int q_{n}(v_{n},v_{-n})f(v_{-n})dv_{-n}.$ Hence, the quantile
allocation is: 
\begin{equation*}
X(t)=q_{n}(V(t)).
\end{equation*}%
In a symmetric auction, a non-decreasing allocation rule can be implemented
if and only if 
\begin{equation*}
X(t)\prec _{w}t^{N-1},
\end{equation*}%
which is sometimes referred to as Border's conditions (see \cite{bord91}).
This argument was developed in detail and also extended to general rank
auctions in \cite{klms21}. \ Thus with $Q\left( t\right) =t^{{N-1}}$, we can
interpret the mechanism design problem as finding the allocation that
maximizes the revenue in a symmetric auction. \ 

\subsection{Revenue-Maximizing Mechanism}

We now fix the information structure to be complete disclosure-- that is, we
impose $W(t)=V(t)$-- and find the optimal mechanism. We can express the
problem of finding the optimal mechanism as solving 
\begin{equation*}
\max_{X\prec _{w}Q}R(V,X).
\end{equation*}%
This is the problem studied by \cite{lomu22} who characterize the optimal
selling policy for a distribution of (discrete)\ qualities to a continuum of
buyers.

We define a function $r_{W}\left( t\right) $ as follows: 
\begin{equation*}
{r}_{W}(t)\triangleq W(t)(1-t),
\end{equation*}%
and write the revenue using the earlier expression (\ref{eq:rev}) as
follows: 
\begin{equation}
R(W,X)=X(0)r_{W}(0)+\int_{0}^{1}r_{W}(t)dX(t).  \label{fdc}
\end{equation}%
We refer to $r_{W}$ as the \emph{pointwise revenue function}. If the seller
makes a take-it-or-leave-it offer that is accepted by $(1-t)$ fraction of
the consumers then the price must be $p=W(t)$, and thus the expected revenue
is $r_{W}(t)$. If we interpret the distribution as a continuum of potential
buyers, then $(1-t)$ is the quantity sold by the seller (the demand) and $%
V(t)$ is the inverse demand function. We thus have a linear optimization
problem in $dX\left( t\right) $ where the weights are given by the pointwise
revenue function.

We define the quantile that maximizes the revenue function: 
\begin{equation*}
t_{m}\triangleq \underset{t\in \lbrack 0,1]}{\arg \max }\ r_{W}(t),
\end{equation*}%
which we assume is uniquely defined. Hence, a monopolist facing a demand $%
V(t)$ would serve a fraction $(1-t_{m})$ of the buyers.


\begin{figure}[h!]
	\centering
	\includegraphics[width=4.3068in,height=2.5365in,keepaspectratio]{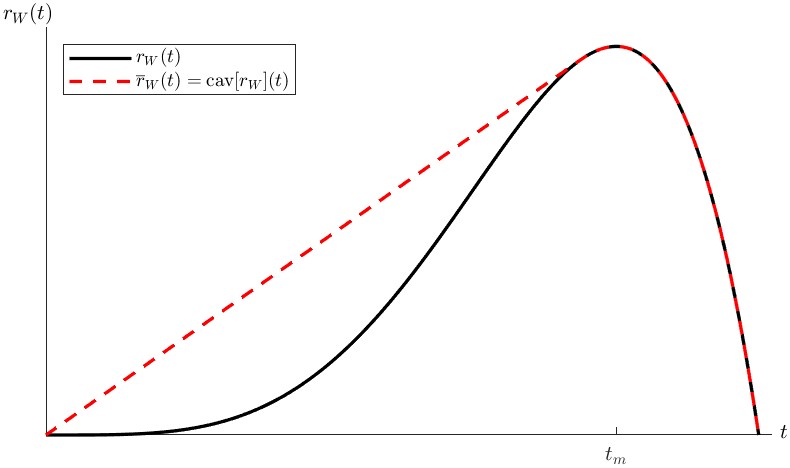}
	\caption{Revenue function $r_{W}(t)$ and its concavification $\overline{r}_{W}(t)$ when $W(t)=t^{4}$.}
	\label{fig1}
\end{figure}

We denote by $\overline{r}_{W}$ the concavification of $r_{W}$; the
concavification of function $r_{W}$\ is the smallest concave function that
is pointwise larger than $r_{W}$, which we illustrate in Figure \ref{fig1}.

To describe the optimal mechanism, it is useful to introduce the notation
necessary to describe the concavification. There exists a collection of
quantiles (and associated open intervals)$\ \left\{ (\underline{t}_{i},%
\overline{t}_{i})\right\} _{i\in I}$ indexed by $i\in I$, with $\underline{t}%
_{i},\overline{t}_{i}\in \lbrack 0,1]$ and 
\begin{equation}
\underline{t}_{i}<\overline{t}_{i}\leq \underline{t}_{i+1},  \label{eq:ti}
\end{equation}%
such that: 
\begin{equation*}
\overline{r}_{W}(t)%
\begin{cases}
=r_{W}(t), & \text{if }t\notin \bigcup\limits_{i\in I}(\underline{t}_{i},%
\overline{t}_{i}); \\ 
>r_{W}(t), & \text{if }t\in (\underline{t}_{i},\overline{t}_{i}).%
\end{cases}%
\end{equation*}%
Hence, the quantiles $(\underline{t}_{i},\overline{t}_{i})_{i\in I}$ define
where $r_{W}$ lies below its concavification. Thus, we can describe the
optimal mechanism.

\begin{theorem}[Optimal Mechanism via Concavification]
\label{dcxs}\quad \newline
An optimal mechanism is given by: 
\begin{equation*}
X(t)=%
\begin{cases}
0, & \text{if }t<t_{m}; \\ 
Q(t), & \text{if }t\geq t_{m}\text{ and }t\notin \bigcup\limits_{i\in I}(%
\underline{t}_{i},\overline{t}_{i}); \\ 
\frac{\int_{\underline{t}_{i}}^{\overline{t}_{i}}Q(t)dt}{\overline{t}_{i}-%
\underline{t}_{i}}, & \text{if }t\geq t_{m}\text{ and }t\in (\underline{t}%
_{i},\overline{t}_{i}).%
\end{cases}%
\end{equation*}%
The revenue generated by the optimal mechanism is: 
\begin{equation}
\max_{X\prec _{w}Q}R(W,X)=\overline{r}_{W}(t_{m})Q(t_{m})+\int_{t_{m}}^{1}%
\overline{r}_{W}(t)dQ(t).  \label{cdc}
\end{equation}
\end{theorem}

That is, the mechanism excludes buyers with quantiles $t<t_{m}$ (using a
reserve price $p=V(t_{m})$) and pools all quantiles (and values) $t>t_{m}$
whenever $r_{W}(t)<\overline{r}_{W}(t)$. By contrast all quantiles $t>t_{m}$
and outside the pooling interval are allocated the efficient allocation:\ $%
X\left( t\right) =Q\left( t\right) $. We provide a proof of the result in
the Appendix that uses majorization (as explained later, similar results
have appeared in the literature and have employed a variety of different
approaches). If $\overline{r}_{W}(t)$ is strictly concave at any 
\begin{equation*}
t\notin \bigcup\limits_{i\in I}(\underline{t}_{i},\overline{t}_{i}),
\end{equation*}%
then the mechanism we provide is the unique optimal mechanism; furthermore,
this condition is satisfied \textquotedblleft generically\textquotedblright\
(where we add the quotation marks to emphasize we have not formalized this).

This theorem characterizes the revenue obtained by the seller as well as the
optimal mechanism that attains this revenue. We first note that the optimal
revenue is obtained using expression \eqref{fdc} with three replacements. We
can replace $r_{W}$ with its concavification (which by construction is
always weakly higher), we replace $X(t)$ with the exogenous inventory $%
Q\left( t\right) $, and the integral with the initial condition are
evaluated at $t_{m}$ instead of zero. The optimal mechanism excludes buyers
below the revenue-maximizing quantile $t_{m}$, pools the allocation of all
quantiles where $r_{W}(t)<\overline{r}_{W}(t)$ and buyers with quantiles $t$
such that $r_{W}(t)=\overline{r}_{W}(t)$ receive their corresponding
efficient allocation.


\subsection{Single Unit Auction}

When we interpret our model as an auction, the specific functional form $%
Q(t)=t^{N-1}$ does not play a role in determining the quantiles of goods
that need to be pooled. As expressed in Theorem \ref{dcxs}, this only
depends on the distribution of values. Once the correct distribution of
quantiles that need to be pooled is determined using the distribution of
values, the optimal auction can be implemented as a direct mechanism where a
buyer whose value is in quantile $t$ receives the good with probability $%
X(t) $. In auctions, there are usually indirect implementations of interest
(such as a first-- or second-- price auction), but we will not delve into
this as it is outside the scope of our survey.

\subsection{Example: Nonlinear Pricing and Single Unit Auction}

\label{examp} 
\begin{figure}[h!]
	\centering
	\includegraphics[width=4.8386in,height=2.5374in,keepaspectratio]{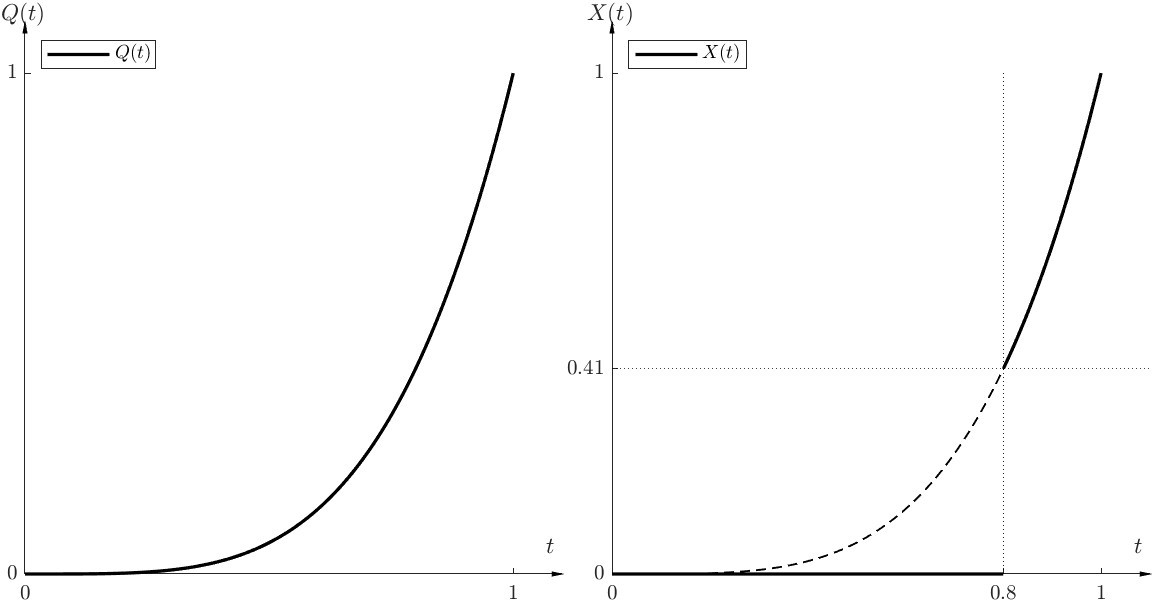}
	\caption{Allocation in the Optimal Auction: Feasible Allocation $Q(t)$ and Optimal Allocation $X(t)$, with $Q \succ_{w} X$.}
	\label{fig1a}
\end{figure}

We illustrate the result with our leading example. If $W\left(
t\right) =V\left( t\right) ={t}^{4}$, then $r_{W}$ is plotted in Figure \ref%
{fig1}. We can see that $r_{W}(t)<\overline{r}_{W}(t)$ in a range of
quantiles $t\in \lbrack 0,3/4]$. However, the revenue-maximizing quantile is 
$t_{m}=4/5$, so all values that would be pooled are, anyway, excluded as
they are below the revenue maximizing quantile $t_{m}$. The unique
seller-optimal allocation is: 
\begin{equation*}
X(t)=%
\begin{cases}
0, & \text{if }t<\frac{4}{5}; \\ 
Q(t), & \text{if }t\geq \frac{4}{5}.%
\end{cases}%
\end{equation*}%
Hence, the seller discards a fraction $4/5$ of the goods and assigns the
remaining ones assortatively. In an auction, this allocation can be
implemented using a second-price auction with a reserve price equal to $%
V(t_{m})$. We present the resulting allocation in Figure \ref{fig1a}.

\subsection{Regular Distributions}

\cite{myer81} identified a regularity assumption that is sufficient for no
pooling, namely monotonicity of the virtual value in the value $v$:%
\begin{equation*}
v-\frac{1-F\left( v\right) }{f\left( v\right) }\text{.}
\end{equation*}%
We thus focus on when the virtual value $\phi \left( t\right) $, stated in
terms of the quantile value function with 
\begin{equation}
\phi (t)\triangleq W(t)-W^{\prime }(t)(1-t)  \label{cxs}
\end{equation}%
is increasing in $t$. This is equivalent to requiring that the revenue
function $W(t)(1-t)$ is concave in $t$.

\begin{corollary}[Monotone Virtual Values]
\label{prop:mon}\quad \newline
If the virtual value $\phi \left( t\right) $ is increasing in $t$, then the
optimal allocation is efficient for all quantiles $t\geq t_{m}$--that is $%
X(t)=Q(t)$-- and excludes quantiles $t<t_{m}$--that is, $X(t)=0$.
\end{corollary}

This proposition provides sufficient conditions so that the optimal selling
mechanism consists of discarding a fraction of the goods and assigning the
remaining ones assortatively. This is the same classic condition that is
traditionally given so that the optimal auction can be implemented as a
second-price auction with a reserve price. Our main example does not satisfy
the condition in this proposition as the virtual utility is first decreasing
and then increasing. But the optimal auction is still a second-price auction
with a reserve price. As explained above, this is because the section where
the allocation would be pooled corresponds to buyers that are excluded
anyway as the virtual utility is indeed negative.

\subsection{Ironing and Feasibility in Mechanism Design}

Our analysis most directly combines two recent papers. The model with a
continuum of buyers and an exogenous inventory of goods is essentially
equivalent to that of \cite{lomu22}, except for the fact that we consider a
continuum of different qualities. However, to analyze the problem, we follow 
\cite{klms21} in expressing the feasibility constraint as a majorization
constraint and then using their characterization of maximization of linear
functionals subject to majorization constraints. We also follow \cite{klms21}
in mapping the model with a continuum of buyers of \cite{lomu22} to optimal
symmetric auctions (\cite{myer81}).

The general problem of maximizing a function subject to a monotonicity
constraint has appeared in various environments. Early on, \cite{muro78}
analyze the problem of a monopolist that can engage in second-degree price
discrimination and provides a solution when monotonicity constraints binds.
In their model, the quality provision is costly, and so this corresponds to
an endogenous inventory, which we do not address in this survey. The seminal
work of \cite{myer81} also provides a solution to the maximization problem
when the monotonicity constraint binds, but in the context of auctions. As
we explained earlier, this can be mapped to our model with an exogenous
inventory when we restrict attention to symmetric auctions, for a more
general analysis of ironing see \cite{toik11} and \cite{nosa07}.

Now, a version of the geometric analysis we have provided here already
appears in \cite{myer81}, however the interpretation of the concavification
(convexification) method appears later on. The connection between the
virtual values and the marginal revenue of a monopolist appears in \cite%
{bukl96}; the connection between the ironing procedure and the
concavification of the revenue function appears in \cite{lomu22}. And,
earlier versions of \cite{lomu22} showed that the same concavification
allows finding the optimal mechanism when the cost is endogenous. Hence,
when the distribution of qualities is endogenous, similar geometric
techniques can be applied (even though there is no relevant feasibility
constraint for the inventory of goods).

The description we have provided is not meant to be exhaustive, but simply
to sample some of the literature to illustrate the different variations and
interpretations of the problem we study that have arisen in the context of
optimal mechanism design.\footnote{%
Similar ironing techniques have also appeared in \cite{akdk21}, \cite{akdk24}%
, \cite{kocw25} and \cite{kowo25} among others.} Our focus on exogenous
inventory allows us to bring together a couple of different interpretations
and use the language of majorization to describe both the feasible and
optimal distributions of qualities. Most importantly, we show that this
framework allows us to closely connect the analysis to the seemingly
unrelated literature on information design, which we analyze next.

\section{Information Design\label{sec:infode}}

We now turn to the information design problem, fixing the mechanism and
optimizing over the information structure. Analogously to our previous
section, our results apply to any distribution of qualities $X$, but the
results can be most naturally interpreted when the mechanism is efficient.
That is, when $X(t)=Q(t)$.

We begin by establishing the distribution of expected values that can be
induced by some information structure, which follows from Blackwell's
theorem. The seller's problem again becomes a linear optimization over the
quantile value function $W$. We show that the optimal information structure
is obtained by concavifying an appropriately constructed function. We then
examine a broad class of distributions of qualities where the inventory of
qualities exhibits a monotone hazard rate. In this case, the analysis yields
a simple dichotomy: a decreasing hazard rate of the quality distribution
implies complete disclosure of information, while increasing hazard rate
implies complete pooling in a single interval, thus no disclosure. The
following subsection connects these results to efficient auctions, showing
that the optimal information structure depends only on the number of
bidders, and then illustrates the results with an example. In the context of
auctions, the corresponding distribution of qualities does not exhibit a
monotone hazard rate, so we get that partial disclosure is optimal. The
final subsection interprets the problem as a form of nonlinear Bayesian
persuasion, closely related to Myerson's ironing argument and to persuasion
problems with \cite{yaar87}-type preferences.

\subsection{Feasibility of Information Structure}

The buyer information structure is summarized by the distribution of values $%
F_{v}$. By \cite{blac51}, Theorem 5, there exists an information structure
that induces a distribution $F_{w}$ of expected values if and only if $F_{w}$
is a mean-preserving contraction of $F_{v}$, that is, 
\begin{equation}
\int_{v}^{\overline{v}}F_{v}({x})dx\leq \int_{v}^{\overline{v}\ }F_{w}({x})dx%
\text{, }\forall {v}\in \lbrack \underline{v},\overline{v}],  \label{eq:majo}
\end{equation}%
with equality for $v=\underline{v}$. \ If $F_{w}~$is a mean-preserving
contraction of $F_{v}$ (or $F_{w}$ majorizes $F_{v}$), we write $F_{v}\prec
F_{w}$. Following \cite{shsh07}, Section 3.A, we have that $F_{w}$ majorizes 
$F_{v}$ if and only if the ex ante quantile value function $V\left( t\right) 
$ majorizes the quantile (expected) value function $W\left( t\right) $: 
\begin{equation*}
W\prec V.
\end{equation*}%
We thus obtain a direct constraint on the quantile functions. 
%

\subsection{Revenue-Maximizing Information Structure}

We now consider the problem of finding the information structure that
maximizes revenue assuming that the mechanism is implementing the efficient
outcome. We can express this problem as follows: 
\begin{equation}
\max_{W\prec V}R(W,Q).  \label{eq:rwq}
\end{equation}%
Here we evaluate the second argument of the revenue $X$ at $X=Q$ because we
are considering an efficient mechanism.

To describe the revenue-maximizing information structure, we define: 
\begin{equation}
{e}_{X}(t)\triangleq 
\begin{cases}
\int_{t}^{1}(1-s)dX(s), & \text{if }t>0; \\ 
\int_{t}^{1}(1-s)dX(s)+X(0), & \text{if }t=0.%
\end{cases}
\label{cdxz}
\end{equation}%
Here, the definition is discontinuous at $t=0$ to account for the rents
extracted from the minimum quality that is offered to all buyers. For now we
omit the discontinuity at $t=0$ to provide an interpretation. Integrating by
parts, we get that for all $t>0$: 
\begin{equation*}
e_{X}(t)=\int_{t}^{1}\left( X(s)-X(t)\right) ds.
\end{equation*}%
This is a measure of the excess quality in the inventory at $t$. In
particular, it is the total additional quality that can be offered to
quantiles above $t.$

This is related to a well-known stochastic order: the excess wealth order
(see Section 3.C in \cite{shsh07}). Say we have two inventories $X,\widehat{X%
}$, then $\widehat{X}$ is greater than $X$ in the excess wealth order if and
only if $e_{X}(t)\leq e_{\widehat{X}}(t)$ for all $t$. The excess wealth
order is a measure of variability. It is precisely the stochastic order that
determines when one inventory leads to more informational rents for every
distribution of expected values. With the notion of excess quality, we can
rewrite the revenue (\ref{eq:re}) as follows: 
\begin{equation}
R(W,X)=e_{X}(0)W(0)+\int_{0}^{1}e_{X}(t)dW(t).  \label{dczc}
\end{equation}%
Here we obtained the same expression as in \eqref{fdc}, but we replaced the
revenue function $r_{W}\left( t\right) $ with the excess quality function, $%
e_{X}\left( t\right) $, and we maximize with respect to the value quantile
function $W\left( t\right) $ rather than the allocation quantile $X\left(
t\right) $. Thus, as we solve for the optimal information design, we permute
the integrator from $dX\left( t\right) $ to $dW\left( t\right) $, and
conversely permute the integrand from $r_{W}\left( t\right) $ to $%
e_{X}\left( t\right) $ (as displayed in (\ref{dczc}) relative to the earlier
problem (\ref{fdc}) in mechanism design.

We can interpret the new expression (\ref{dczc}) as follows. The seller
sells all goods at a baseline price $W\left( 0\right) $, which generates a
baseline revenue $W\left( 0\right) e_{X}\left( 0\right) $, and then adds a
surcharge. The surcharge consists of an additional price increment $dW\left(
t\right) $ per unit of quality to all goods of quality higher than $Q\left(
t\right) $. Hence, the price increment $dW\left( t\right) $ generates a
revenue $e_{X}\left( t\right) $, which is the total excess quality relative
to $Q\left( t\right) $. Thus, the linear weight of the maximization problem
are now the excess qualities $e_{X}\left( t\right) $.

\begin{figure}[h!]
	\centering
	\includegraphics[width=4.3171in,height=2.5374in,keepaspectratio]{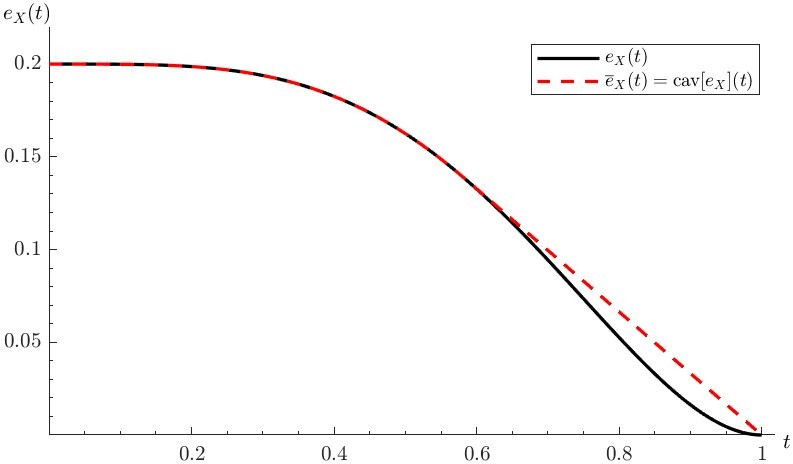}
	\caption{The excess function $e_{X}$ and its concavification $\overline{e}_{X}$ when $X(t)=t^{4}$.}
	\label{fig2}
\end{figure}

Additionally, we also have that the maximization problem (\ref{eq:rwq})\ is
subject to a strong majorization constraint. In Figure \ref{fig2} we plot $%
e_{X}(t)$ for our canonical example where $X(t)=t^{4}$. Although the
quantile distributions of $W\left( t\right) $ and $X\left( t\right) $\ are
symmetric, the linear coefficients in the integrals, $r_{W}(t)$ and $%
e_{X}(t) $, respectively are not the same.

To obtain an intuition for why we have a discontinuity in the definition of %
\eqref{cdxz} suppose that the seller offers a homogeneous quality, that is, $%
X(t)=1$. We then have that $dX(t)=0$ for all quantiles $t\in (0,1)$, and so
also $e_{X}(0)=0$ in this interval. However, in this case, the revenue
obtained is exactly the quality $X(0)$ times the minimum value $W(0)$.
Hence, the discontinuity allows us to correctly account for the fact that
the optimal mechanism extracts $X(0)W(0)$ from all buyers, and then the
seller extracts extra revenue from each quality increment $dX(t)$.

We can thus describe the optimal information structure as follows. We denote
by $\overline{e}_{X}\left( t\right) $ the concavification of {$e$}$_{X}$
(smallest concave function that is pointwise larger than the function
itself). To describe the optimal mechanism it is useful to introduce
notation necessary to describe the concavification. As before in the
description of the mechanism design problem (see (\ref{eq:ti})), there
exists a sequence of quantiles $\left\{ \underline{t}_{i},\overline{t}%
_{i}\right\} _{i\in I}$, with $\underline{t}_{i},\overline{t}_{i}\in \lbrack
0,1]$ and 
\begin{equation*}
\underline{t}_{i}<\overline{t}_{i}\leq \underline{t}_{i+1},
\end{equation*}%
such that: 
\begin{equation}
\overline{e}_{X}(t)\text{\ }%
\begin{cases}
=e_{X}(t), & \text{if }t\notin \bigcup\limits_{i\in I}(\underline{t}_{i},%
\overline{t}_{i}); \\ 
>e_{X}(t), & \text{if }t\in (\underline{t}_{i},\overline{t}_{i}).%
\end{cases}
\label{eq:ex}
\end{equation}%
Hence, the quantiles $(\underline{t}_{i},\overline{t}_{i})_{i\in I}$ define
where $e_{X}$ lies below its concavification. We can thus describe the
optimal information structure.

\begin{theorem}[Optimal Information Structure via Concavification]
\label{dcxs2}\quad \newline
The optimal information structure generates revenue: 
\begin{equation}
\max_{W\prec V}R(W,X)=\overline{e}_{X}(0)V(0)+\int_{0}^{1}\overline{e}%
_{X}(t)dV(t).  \label{cdxaa}
\end{equation}%
The optimal information structure is given by: 
\begin{equation*}
W(t)=%
\begin{cases}
V\left( t\right) , & \text{if }t\notin \bigcup\limits_{i\in I}(\underline{t}%
_{i},\overline{t}_{i}); \\ 
\frac{\int_{\underline{t}_{i}}^{\overline{t}_{i}}V(s)ds}{\overline{t}_{i}-%
\underline{t}_{i}}, & \text{if }t\in (\underline{t}_{i},\overline{t}_{i}).%
\end{cases}%
\end{equation*}
\end{theorem}

That is, the optimal information structure pool values whenever {$e$}$%
_{X}(t)<\overline{e}_{X}(t)$.\ This theorem characterizes the revenue
obtained by the seller as well as the optimal information structure that
attains this revenue. We first note that the optimal revenue is obtained
using expression \eqref{dczc} but we replaced $e_{X}$ with its
concavification (which by construction is always weakly higher), and we
replaced the distribution of expected values $W(t)$ with the exogenous
distribution of values $V(t)$. The optimal information structure pools the
values of all quantiles where $e_{X}(t)<\overline{e}_{X}(t)$ and all
quantiles such that $e_{X}(t)=\overline{e}_{X}(t)$ learn their value. We
provide a proof of the result in the Appendix, where, as in Theorem \ref%
{dcxs}, we use results and ideas from majorization theory.

Also, as before, if $\overline{e}_{X}(t)$ is strictly concave at any 
\begin{equation*}
t\notin \bigcup\limits_{i\in I}(\underline{t}_{i},\overline{t}_{i}),
\end{equation*}%
then the mechanism we provide is the unique optimal information structure;
furthermore, this condition is satisfied \textquotedblleft
generically\textquotedblright\ (where we add the quotation marks to
emphasize we have not formalized this).

\subsection{Inventories with Monotone Hazard Rate\label{subsec:mhr}}

We now consider the problem of finding the revenue-maximizing information
structure when the mechanism is efficient. Hence, we replace $X(t)=Q(t)$ and
proceed to find the solution as a function of $Q(t)$. For simplicity, we
assume that $Q(0)=0$.

We can infer the solution to \eqref{cdxaa} by looking at the curvature of {$%
e $}$_{Q}$. If {$e$}$_{Q}$ is concave, then the solution is full disclosure.
If {$e$}$_{Q}$ is convex, then the solution is no disclosure. If {$e$}$_{Q}$
is linear, then any information structure is a solution. We can then provide
conditions for {$e$}$_{Q}$ to be convex or concave.

If $Q^{\prime }(t)(1-t)$ is decreasing, then the optimal information
structure displays no disclosure. If $Q^{\prime }(t)(1-t)$ is increasing,
then the optimal information structure is to provide full disclosure of
values. We can express the monotonicity condition regarding $Q^{\prime
}(t)(1-t)$ in terms of the hazard rate: 
\begin{equation*}
Q^{\prime }(s)(1-s)=\frac{1-F_{q}(q)}{F_{q}^{\prime }(q)}.
\end{equation*}%
Hence, this corresponds to the inverse hazard rate of the distribution of
qualities.

\begin{proposition}[Optimal Disclosure]
\label{exinv2}\quad \newline
If the distribution of qualities has an increasing hazard rate, then the
optimal information structure is no disclosure. If the distribution of
qualities has a decreasing hazard rate, then the optimal information
structure is full disclosure.
\end{proposition}

We thus have that the optimal information structure critically depends on
the hazard rate of the distribution of qualities. The reason is that the
allocation is efficient $X(t)=Q(t)$ so revenue is determined by competition
from the nearby value. For example, if the qualities are all the same, say $%
Q(t)=1$, then the revenue is simply the lowest value, so full pooling is
optimal. On the other extreme, if the distribution of qualities is Pareto,
say $Q(t)=(1-t)^{-\alpha }$ with $\alpha >1$, then the surplus gains from
allocating the good efficiently outweigh the informational rents that are
generated.


\subsection{Single Unit Auction\label{auctions}}

As we noted earlier, when $Q(t)=t^{N-1}$ the model can be interpreted as an
auction with $N$ bidders. Furthermore, since the allocation is the efficient
one, in the context of an auction, this allocation can be implemented with a
second-price auction without a reserve price. We thus recover the analysis
of \cite{bhms22}. We will also inherit from Theorem \ref{dcxs2} the feature
that the optimal information structure depends only on quantiles and not on
the distribution of values. In contrast to the optimal mechanism setting, we
now have that the quantiles $Q$ determine the quantiles that need to be
pooled. Hence, under the interpretation of an auction environment, the
revenue-maximizing information structure will depend only on the number of
bidders. The optimal information structure will not be either of the cases
discussed in Proposition \ref{exinv2}, as the hazard rate associated with $%
Q(t)=t^{N-1}\ $is not monotone. We illustrate this in the following section
with our main example.

\subsection{Example: Information Design in Second Price Auction\label%
{examp:inf-rev}}

We now illustrate the result of the optimal information structure for the
case of an exogenous inventory given by the quantile allocation function $%
Q\left( t\right) =t^{4}$. \ We present our results with the auction
interpretation as in \cite{bhms22}, but the reader should keep in mind that
the results also apply to the screening problem with exogenous inventory. \ 

We characterize the structure of the optimal information policy. For this,
we first plot $e_{Q}$ in Figure \ref{fig2}. We can see that it is concave at
low quantiles and convex at high quantiles. We then have that the optimal
information structure consists of pooling high values and providing full
disclosure to low values, as suggested by the concavification in Figure \ref%
{fig2}. We thus have that in our main example, the optimal information
structure is: 
\begin{equation}
W(t)=%
\begin{cases}
V(t), & \text{if }t<t^{\ast }; \\ 
\frac{\int_{t^{\ast }}^{1}V(s)ds}{1-t^{\ast }}, & \text{if }t\geq t^{\ast };%
\end{cases}
\label{eq:opt-info}
\end{equation}%
where the optimal threshold $t^{\ast }\approx 0.6$ is the solution to: 
\begin{equation*}
e_{X}^{\prime }(t^{\ast })(1-t^{\ast })=e_{X}(1)-e_{X}(t^{\ast }).
\end{equation*}%
Hence, this threshold is just the threshold that determines the
concavification of $e_{X}$.

Intuitively, the optimal structure reveals the exact value of all bidders
whose values fall below a fixed threshold, determined by the quantile $%
t^{\ast }$. For values above this threshold, the mechanism reveals only the
information that the value lies above the cutoff, effectively pooling these
types at a common posterior mean. This amounts to an \textquotedblleft upper
censorship\textquotedblright\ of the value distribution.
\begin{figure}[h!]
	\centering
	\includegraphics[width=4.8386in,height=2.5374in,keepaspectratio]{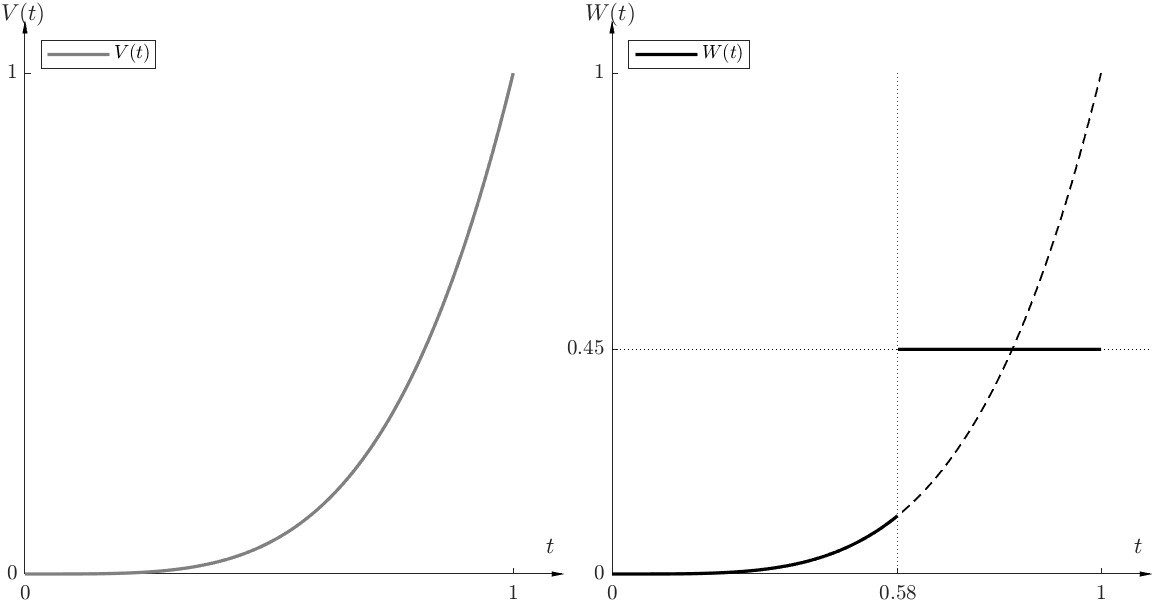}
	\caption{Optimal Information Design with $V(t)$ and $W(t)$, where $V(t) \succ W(t)$.}
	\label{fig2a}
\end{figure}

The role of this information structure is to sustain competition at the
upper tail of the distribution, at the cost of efficiency. In particular,
all values above the threshold $F^{-1}(q^{\ast })$ are bundled together into
a single mass point. In doing so, the seller depresses the information rent
of the winning bidder, thereby raising expected revenue.

\subsection{Non-Linear Bayesian Persuasion\label{non-linear}}

In the classic Bayesian persuasion problem of \cite{kage11}, a sender
chooses an information structure - a distribution of posteriors - to
maximize her expectation of her indirect utility from the action of a
receiver informed of the posterior. Because the sender is an expected
utility maximizer, the objective is linear in the indirect utility of the
sender. Several papers have examined the case where the receiver only cares
about their expectation of the state, e.g., \cite{kolo18} and \cite{dwma19}.
\ In this case, all the sender cares about is the distribution over
expectations of the state that can be induced by an information structure.
The set of such distributions is the set of mean-preserving contractions of
the ex ante distribution of expected values. In the language of
majorization, it is the set of distributions that majorize the ex ante
distribution of values. \ \cite{klms21} unify and generalize these results
in their Theorem 2, which characterizes the extreme points of the set of
mean preserving contractions of a distribution. \ 

The information design problem that we study in this section does not belong
to this class. We analyze the problem in quantile space, working with the
inverse of the distribution over values. We observed that the seller's
revenue is linear in the quantile value function and the seller is choosing
among quantile value functions that are majorized by the ex ante quantile
value function. Thus, we appeal to Theorem 1 of \cite{klms21} which
characterizes extreme points of the mean preserving spreads of a
distribution. Note that \cite{myer81} already characterized optimal auctions
using quantile value functions, which is why the information design in this
section essentially follows \cite{myer81}. \ \ 

The information design in this section is equivalent to a non-linear
Bayesian persuasion problem, i.e., one where the sender's preferences are
not expected utility. \cite{yaar87} considered a dual theory of choice under
risk where the decision maker's preferences were linear in lottery payments
but non-linear in probabilities. \ Our problem is equivalent to a Bayesian
persuasion problem with Yaari preferences.\footnote{%
We are grateful to Roberto Corrao for pointing this out to us. \ } \cite%
{jequ23} analyzes this problem and shows its relevant for a number of
applications (with expected utility maximizers). \cite{gmsz21} consider an
auction model with endogeneous value that leads to a reduced-from model
where agents have nonexpected utility with Yaari preferences. \cite{gmsz22}
investigate the optimal auction for non-expected utility models which
include the dual preferences of Yaari.

\cite{kowo25} show that the problem of gerrymandering congressional
districts corresponds to a linear Bayesian persuasion problem when the
gerrymanderer cares about maximizing the expected number of seats. \ An
early version of their paper (\cite{kowo20}) showed that one could solve the
problem if the gerrymanderer cares about a non-linear function of the number
of seats in a special case where their aggregate shock had a uniform
distribution. \ They solved the problem adapting Myerson's ironing result
and this result too fits our setting. \ 

Also related are a number of papers where one assumes that the state is a
real number and information is restricted to take the form of a partition,
so that signals are a monotonic deterministic function of the state: \cite%
{rayo13} and \cite{onra23}. \ 

%
%
%
%

\section{Combining Mechanism Design and Information Design\label{sec:combine}%
}

We now combine mechanism design and information design by allowing the
seller to choose both the allocation and the information structure.

We first define the problem and discuss its interpretation. We then show
that optimal mechanisms can be represented by monotone partitional
structures, intervals in which types are either fully separated or pooled,
with common support between the distributions of expected values and
qualities. The main result establishes that pooling is always optimal:
whenever both information and allocation are jointly optimized, every
interval in the quantile space involves pooling. We then revisit the
interpretation of the auction and provide the solution to our leading
parametrization. We conclude by arguing that the main takeaway-- that is,
that pooling is optimal-- is robust and provide alternative ways to argue
for this.

\subsection{Maximizing Over Information and Allocation}

Suppose that the seller can choose both the allocation and the information
structure that are chosen to maximize revenue. The joint optimization
problem can be succinctly expressed as solving: 
\begin{equation}
\max_{\substack{ W\prec V  \\ X\prec _{w}Q}}R(W,X).  \label{swp}
\end{equation}%
We thus obtain a maximization problem over two quantile functions subject to
two majorization constraints. We now refer to a \emph{joint mechanism} as an
information structure $W$\ and an allocation $X$: 
\begin{equation*}
M\triangleq (W,X).
\end{equation*}%
Here there is some abuse of notation because in previous sections a
mechanism only referred to and allocations and associated transfers. But,
since the seller can now also control the information disclosure, we augment
the concept of a mechanism to incorporate both dimensions of the problem.

Before we proceed it is useful to briefly discuss the implementation of a
mechanism. One way to think of a mechanism is that the seller first designs
the information structure and then, given this fixed information structure,
design an optimal mechanism. Under this interpretation it is without loss of
generality to assume that the mechanism is a direct mechanism. An
alternative interpretation is that the seller first chooses a price menu $%
T:q\rightarrow \mathbb{R}_{+}$ specifying the price at which each good is
sold and then, given this menu, provides some information to buyers about
their own value. Under this alternative interpretation it is without loss of
generality to think of the information as being a recommendation of which
quality should the buyer purchase. Furthermore, this opens the door to
different interpretations of why the seller might be able to design the
information without being able to engage in explicit discrimination. For the
purpose of this survey we will simply examine the reduced-form description %
\eqref{swp}; a more detailed discussion of the interpretation and
implementation is given in \cite{behm26}.

Our main analysis proceeds in two steps. We first leverage the analysis we
have carried out so far to provide general properties of the optimal
mechanism. We then provide the main qualitative property of the solution
which is that there is always pooling. We then interpret the result in the
context of auctions and provide an illustrative example. Finally, we
conclude this section by explaining in what sense the pooling that arises in
this problem is more pervasive and sustained than the pooling that arises
when we maximize over distribution of values or allocations, but not both.
The results we provide here can be found in the published version of \cite%
{behm26}. Here we follow more closely the exposition found in an earlier
working paper version \cite{behm24}.\footnote{%
The published version allowed for a general class of production functions,
so the distribution of qualities is allowed to be endogeneous. In the
earlier working paper we focused on the case in which the seller had an
exogeneous inventory.}

\subsection{Structure of the Optimal Mechanism\label{sec:om}}

We now provide a characterization of the optimal mechanism. \ The mechanism
specifies a distribution of expected values $W$ as the choice of information
and a distribution of (expected) qualities $X$ as the choice of available
qualities. The bilinear structure allows us to use Propositions \ref{dcxs}-%
\ref{dcxs2} to obtain that the quantile functions of expected values and
qualities have monotone partitional structure. Furthermore, a simple
argument allows us to show that both quantile functions must have the same
support. The common support can be obtained from using the revelation
principle and the taxation principle simultaneously. In other words, it is
without loss of generality to focus on mechanisms in which there is a
bijection between the number of possible expected values and expected
qualities offered to buyers.

We now introduce some language for critical properties of the value and
quality distributions. A quantile function $W$is said to be \emph{monotone
partitional} if $\left[ 0,1\right] $ is partitioned into countable intervals 
$[\underline{t}_{i},\overline{t}_{i})_{i\in I}$ and, writing $I$ for the
labels, we have 
\begin{equation}
W\left( t\right) \triangleq 
\begin{cases}
V\left( t\right) , & \text{ if }t\notin \bigcup\limits_{i\in I}[\underline{t}%
_{i},\overline{t}_{i})\text{ }; \\ 
\frac{\int_{\underline{t}_{i}}^{\overline{t}_{i}}V\left( s\right) ds}{%
\overline{t}_{i}-\underline{t}_{i}}, & \text{ if }t\in \lbrack \underline{t}%
_{i},\overline{t}_{i})\text{ for some }i\in I.%
\end{cases}
\label{eq:ep}
\end{equation}%
and each interval either has pooling, i.e., buyers know only that their
value corresponds to a quantile in the interval $\left( \underline{t}_{i},%
\overline{t}_{i}\right) $ or full disclosure, that is, all buyers with
values corresponding to quantiles not in any interval $[\underline{t}_{i},%
\overline{t}_{i}]$ know their value. The expectation can be written
explicitly in terms of the quantile function as follows: 
\begin{equation*}
w_{i}=\frac{\int_{\underline{t}_{i}}^{\overline{t}_{i}}V\left( t\right) dt}{%
\overline{t}_{i}-\underline{t}_{i}}.
\end{equation*}%
The distribution of expected values $W$ is said to be \emph{monotone pooling}
if all quantiles are pooled, i.e.:%
\begin{equation*}
\bigcup\limits_{i\in I}[\underline{t}_{i},\overline{t}_{i})=\left[ 0,1\right]
\text{.}
\end{equation*}

We can similarly define monotone partitional and monotone pooling
distributions $X$\ of qualities. However, the quality distribution $X$ needs
to be only \emph{weakly} majorized by $Q$. We say $X$ is a \emph{weak}
monotone partitional distribution if there exists a monotone partitional
distribution $\widehat{X}$ and an indicator function $\mathbb{I}_{t\geq
t_{m}}$ such that: 
\begin{equation*}
X(t)=\widehat{X}(t)\cdot \mathbb{I\ }_{t\geq t_{m}}(t),
\end{equation*}%
where $\mathbb{I}_{t\geq t_{m}}(t)$ is the indicator function: 
\begin{equation}
\mathbb{I}_{t\geq t_{m}}(t)\triangleq 
\begin{cases}
1 & \text{if }t\geq t_{m}; \\ 
0 & \text{if }t<t_{m}.%
\end{cases}
\label{eq:ind}
\end{equation}%
In other words, a weak monotone partitional distribution is generated by
first taking all qualities corresponding to quantiles below $t_{m}$ and
reducing these qualities to zero, and then generating a monotone partitional
distribution (where some low qualities have been reduced to 0). A \emph{weak}
monotone pooling distribution is defined in the analogous sense, where all
intervals are pooling.

To simplify the exposition, henceforth we omit the qualifier ``weak''. Thus,
when we say a distribution of qualities $X$ is a monotone partitional (or
monotone pooling) distribution $X$, it is always in the weak sense.

Proposition~1 in \cite{klms21} shows that the set $\{W:W\prec V\}$ is a
compact and convex set, and their Theorem 1 shows that the extreme points of
this set are given by \eqref{eq:ep}. The same analysis applies for $X$, the
only difference is that the set of extreme points of the set $\{X:X\prec
_{w}Q\}$ is the set of weak partitional structures (see Corollary 2 in \cite%
{klms21}). The objective function \eqref{swp} is bilinear in $(W,X)$. Hence,
following Bauer's maximum principle (\cite{baue58}), the maximization
problem attains its maximum at an extreme point of $\{W:W\prec V\}$.

We say that $W$ and $X$ have \emph{common support} if the induced \emph{%
partitions of quantiles} that generate the monotone partition distributions
are the same. We obtain the following result.

\begin{proposition}[Monotone Partitional Distribution]
\label{lem:nec}\quad \newline
There exists a solution $\left( W^{\ast },X^{\ast }\right) $\ to the joint
optimization problem \eqref{swp} that has monotone partitional structure
with common support.
\end{proposition}

Hence, the bilinear structure of the problem guarantees that there exists a
solution in which the quantile distribution of expected values and qualities
are extreme points of their respective feasible sets. The intuition for the
common-support result is that it is (weakly) suboptimal to provide
information if that does not lead to different choices-- which implies that $%
W^{\ast }$ has the same support as $X^{\ast }$-- and it is (weakly)
suboptimal to provide multiple options to buyers with the same expected
value -- which implies that $X^{\ast }$ has the same support as $W^{\ast }$.
In other words, the result is obtained by simultaneously applying the
revelation principle and the taxation principle to the solution of the
problem.

\subsection{Optimality of Pooling\label{sub:optimality}}

We next show that, indeed, every interval is a pooling interval in an
optimal mechanism.

\begin{theorem}[Optimality of Pooling]
\label{thm:pool}\quad \linebreak Every solution $(W^{\ast },X^{\ast })$\ to
the joint optimization \eqref{swp} consists of pooling intervals only.
\end{theorem}

We provide the main arguments of the proof. Consider a mechanism in which
there is full disclosure and full separation in some interval $[t_{1},t_{2}]$
(i.e., $W(t)=V(t)$ and $X(t)=Q(t)$ for all $t\in \lbrack t_{1},t_{2}]$).
Suppose that the seller pools the allocation of all values in an interval $%
[t_{1},t_{2}]$, so that all values get the average quality on this interval.
How much lower would the profit be? We begin the argument with the virtual
values given by: 
\begin{equation}
\phi (t)\triangleq V(t)-V^{\prime }(t)(1-t).  \label{vv}
\end{equation}%
The profit generated is the expectation of the product of the virtual values
and the qualities: 
\begin{equation*}
\Pi \triangleq \int_{0}^{1}Q(t)\phi (t)dt.
\end{equation*}%
We denote the (conditional) mean and variance of virtual values and
qualities in the interval $[t_{1},t_{2}]$\ by:%
\begin{align}
{\mu }_{\phi }\triangleq & \frac{\int_{t_{1}}^{t_{2}}\phi (t)dt}{t_{2}-t_{1}}%
; & \quad {\mu }_{q}\triangleq & \frac{\int_{t_{1}}^{t_{2}}Q(t)dt}{%
t_{2}-t_{1}};  \label{notationmean} \\
\quad {\sigma }_{\phi }^{2}\triangleq & \frac{\int_{t_{1}}^{t_{2}}(\phi (t)-{%
\mu }_{\phi })^{2}dt}{t_{2}-t_{1}}; & \quad {\sigma }_{q}^{2}\triangleq & 
\frac{\int_{t_{1}}^{t_{2}}(Q(t)-{\mu }_{q})^{2}dt}{t_{2}-t_{1}}.
\label{notationvariance}
\end{align}%
As we only compute conditional mean and variance in the interval $%
[t_{1},t_{2}]$, we can safely omit an index referring to the interval $\left[
t_{1},t_{2}\right] $ in the expression of $\mu $ and $\sigma $. The first
step of the proof shows that the revenue losses due to the pooling of
qualities in the interval $[t_{1},t_{2}]$ are bounded by the following: 
\begin{equation*}
{\sigma }_{\phi }{\sigma }_{q}(t_{2}-t_{1}).
\end{equation*}%
Hence, pooling the qualities generates third-order profit losses when the
interval is small (since each of the terms multiplied is small when the
interval is small).

If in addition to pooling the qualities, we pool the values in this
interval, we can reduce the buyers' information rent. When only qualities
are pooled-- but not the values-- then the quality increase that the values
assigned to the pooled quality get relative to the values just below the
pool is the quality difference $\mu _{q}-Q(t_{1})$ priced at $t_{1}$. After
pooling the values, the price of the quality increase is computed using the
expected value conditional on being in this interval: 
\begin{equation*}
{\mu }_{v}\triangleq \mathbb{E}[V\left( t\right) \mid t\in \lbrack
t_{1},t_{2}]].
\end{equation*}%
Hence, pooling the values increases the payments for every quantile value
higher than $t_{1}$ by an amount: 
\begin{equation*}
(\mu _{v}-V(t_{1}))({\mu }_{q}-Q(t_{1}))(1-t_{1}).
\end{equation*}%
Here the first two terms being multiplied are small when the interval is
small. However, payments are marginally increased for all values higher than 
$t_{1}$, which is a non-negligible mass of values (i.e., $(1-t_{1})$ is not
small). In other words, pooling values increases the price of the quality
improvement $(\mu _{q}-X(t_{1}))$ for all values higher than $t_{1}$. Hence,
pooling values generates a second-order benefit which always dominates the
third-order distortions.

\subsection{Example:\ Nonlinear Pricing and Product Design}

\begin{figure}[h!]
	\centering
	\includegraphics[width=4.951in,height=2.5374in,keepaspectratio]{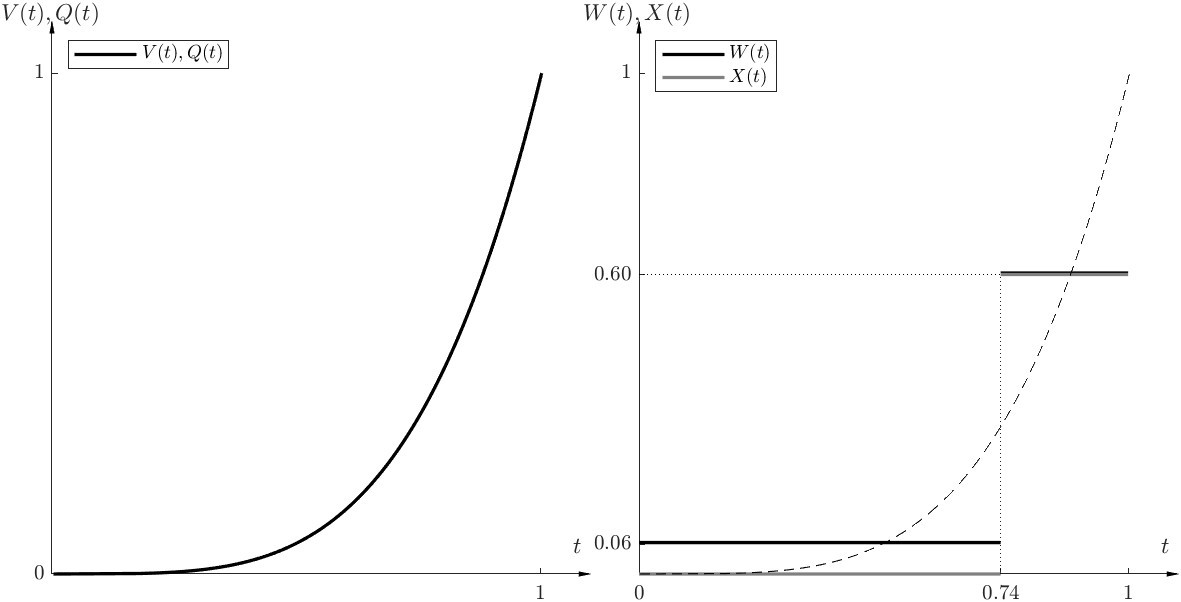}
	\caption{The given distributions of values $V(t)=t^{4}$ and qualities $Q(t)=t^{4}$ are depicted on the left. The associated optimal distributions $W(t)$ and $X(t)$, which are monotone partitional distributions, are depicted on the right.}
	\label{fig3a}
\end{figure}

In Figures \ref{fig3a} we illustrate the optimal mechanism in an example. In
sharp contrast to the earlier analysis of either the optimal allocation or
the optimal information structure, the optimal joint mechanism now leads to
pooled intervals everywhere as established by Theorem \ref{thm:pool}. In the
current example, this leads to there being just two qualities offered and
all values being assigned to one of two pools of values. We can see that the
jumps of $W^\ast$ and $X^\ast$ occur on the same quantiles, as established
in Proposition \ref{lem:nec}. 

\subsection{Auctions}

We can interpret the model as a seller that designs both the information
structure and an optimal symmetric auction. This problem is studied in
greater generality by \cite{bepe07}, where they do not restrict attention to
symmetric auctions. And, in fact, they show that asymmetric information
structures and auctions are optimal. With the restriction to symmetric
outcomes, we then find that with five bidders, there will be two categories
of bidders, those with low and high (expected)\ value.

\subsection{The Robustness of Pooling}

We conclude this section by arguing the pooling we obtain from maximizing
bilinear environments is a more robust feature than the results presented
here. For expositional purposes we analyzed the situation where the seller
has an exogenous inventory, but in the published version of our work we
allow for the inventory to be endogenous (for example, as in \cite{muro78}).
While the result of a monotone partition information structure is based on
the linear utility in $v$, some versions of our main results continue to
hold in a nonlinear utility environment (see Theorem 5 in the working paper
version \cite{behm24}). In the published version, \cite{behm26}, we show
that even with discrete values some pooling is optimal as long as the
entropy of the distribution is not lower than $\log (8)$ (this is the
entropy of a uniform distribution with 8 values). Hence, we get that pooling
arises even in already informationally coarse environments. Overall, we find
that the results pointing towards pooling are of qualitatively different
nature than those presented in Sections \ref{sec:mec}-\ref{sec:infode} when
only one dimension of the design problem was controlled by the seller.

\section{Consumer Surplus\label{sec:cons}}

So far, our analysis has focused on examining how controlling the allocation
and/or information can improve the revenue of the seller. We now examine how
these choices can also be deployed to improve consumer welfare. In fact, we
show that the objective function which represents consumer surplus, as
suggested earlier in the equality \ (\ref{eq:psy}) towards the end of
Section \ref{sec:env} turns out to be the same as those studied in Sections %
\ref{sec:mec}-\ref{sec:infode} after a simple exchange of the quantile
functions. Moreover, the majorization constraints will only be mildly
adjusted relative to those studied in the sections of revenue maximization.
We use these adjustment from weak to strict majorization (and \ conversely)
to emphasize their role in the optimization problems. We conclude this
section by analyzing the case in which both the information and the
mechanism are chosen but the objective is not necessarily revenue
maximization. We argue that the problem becomes more nuanced with a variety
of different problems that can be interpreted differently, while keeping
much of the structure of the original problem.

\subsection{Consumer Surplus}

We begin by highlighting the symmetry between buyers and sellers. Whenever
the boundary condition $W(0)=0$ or $X\left( 0\right) =0$, given earlier in
condition (\ref{eq:bou}) and the distributions are absolutely continuous, we
have that: 
\begin{equation}
R(W,X)=U(X,W).  \label{sym}
\end{equation}%
That is, for a given pair of distributions $(W,X)$ \ the seller obtains the
same revenue as the consumer surplus that the buyers would have obtained if
we had exchanged the roles of the distributions. Thus, the objective are
identical up to an exchange symmetry that requires the swapping of the
quantile function $W$ and $X$. Throughout this section we assume that the
boundary conditions, $W(0)=0$ or $X\left( 0\right) =0,$ hold to simplify the
exposition.

We now briefly provide an interpretation of the symmetry in the objective
functions. The problem of consumer surplus maximization can be found as a
linear optimization, where we only need to modify the weights. Table \ref{T1}
summarize the different objectives with the different possible instruments`
being controlled. 
\begin{table}[h]
\centering\raisebox{-2.0ex}{\shortstack{\textbf{Instrument}}} \hspace{0em} 
\begin{tabular}{c|cc}
& \multicolumn{2}{c}{\textbf{Objective Function}} \\ \cline{2-3}
& \textit{R} & \textit{CS} \\ \hline
Allocation & $(r_{W},\prec _{w})$ & $(e_{W},\prec _{w})$ \\ 
Information & $(e_{X},\prec )$ & $(r_{X},\prec )$ \\ 
&  & 
\end{tabular}%
\caption{Exchange Symmetry of Instruments for Revenue and Consumer Surplus}
\label{T1}
\end{table}

The above table summarizes the relationship between the design optimization
problems and their corresponding representation in terms of linear weights
and type of majorization constraints.%

While the expressions for revenue and consumer surplus are symmetric with
respect to the roles of allocation and information, two aspects of the
problem modify the symmetry in a subtle manner when controlling either
allocation or information. First, so far, we have focused on revenue
maximization, which breaks the symmetry between allocation and information.
In the interpretation of a two-sided matching, controlling the allocation is
the same as pooling sellers, which is the side of the market being
maximized, while controlling the information is pooling buyers, which is the
opposite side of the market being maximized. A second aspect that modify the
symmetry is that, when controlling the allocation, we consider a weak
majorization constraint rather than a strong one. This difference in the
constraint is minor and only results in discarding some goods or,
equivalently, excluding some consumers.

\subsection{Symmetry and Two-Sided Matching}

We have interpreted the model as a seller (monopolist) that offers a variety
of goods to buyers. The transfer payments are then determined by the
allocation and the requirement that the mechanism must be incentive
compatible. However, we could have also interpret the model as consisting of
a continuum of different sellers indexed by $t\in \lbrack 0,1]$, each
offering a good of specific quality $Q(t)$. We would then have a two-sided
matching market and look for the prices that arise in a competitive
equilibrium. In this case, we would have reached the same formulas (see \cite%
{chmn10} or \cite{gali16} for a detailed connection in the context of
matching markets under complete information with quasilinear utility).

We obtain complete symmetry only when $W(0)=0$ or $X(0)=0$. The reason is
that we have assumed that the lowest-value buyer obtains zero surplus. This
condition is implied by the participation constraint when $W(0)=0$ or $%
X(0)=0 $. However, if $W(0), \ X(0)>0$, then the lowest-value buyer obtains
zero surplus only because we assumed that the seller has the all \ the
bargaining power. Hence, when $W(0),\ X(0)>0$, the transfers in a two-sided
matching market are determined by the allocation and the bargaining power,
where the latter determines the surplus of the lowest seller and buyer. The
bargaining power also becomes relevant when we have discontinuities in the
allocation or the information structure, which will become clear as we
provide our analysis. We do not delve into this interpretation any further,
and instead simply leverage the symmetry to study the consumer-optimal
allocation and mechanism.

\subsection{Consumer-Optimal Design\label{examp:inf-cons}}

We now find the consumer-optimal mechanism and the consumer-optimal
information structure.

\paragraph{Consumer-Optimal Allocation}

We begin by considering the consumer-optimal mechanism where the choice
variable is the quantile allocation function $X$: 
\begin{equation}
\max_{X\prec _{w}Q}U(W,X),  \label{cions}
\end{equation}%
where as mentioned we assume that $W(0)=0.$ Appealing to the symmetry
argument in \eqref{sym}, we obtain that analytically this is the same
problem as 
\begin{equation*}
\max_{X\prec Q}R\left( X,W\right) ,
\end{equation*}%
studied in Section \ref{sec:infode}, except for the fact that the
majorization constraint in the consumer surplus maximization problem is weak
while earlier in the revenue maximization problem we had a strict
majorization inequality associated with the Bayesian consistency . However,
it is easy to verify that in the optimum the majorization constraint will be
binding. Formally, this can be proved by noting that $e_{X}(t)$ is
non-increasing. Intuitively, discarding goods is never optimal for the
buyers. Hence, the solution will indeed be the same as the one in Section %
\ref{sec:infode}.

While we analytically obtain the same problem as finding the
revenue-maximizing information, we now have a different interpretation. In
this case, we recover the problem of finding the pooling of qualities that
maximizes buyers' surplus in a two-sided matching market. The problem can be
interpreted as a planner that wants to provide goods to consumers to
maximize their surplus but can only screen buyers using a costly screening
device, like wasteful screening (\cite{klms21} or \cite{cond12}). A general
solution is provided by \cite{klms21} (see Proposition 4 for the result, and
the discussion therein for a broader discussion on contest design). In fact,
in Proposition \ref{exinv2} in Section \ref{subsec:mhr} we recover the same
solution as \cite{klms21} (see Proposition 4.3). They show that the
consumer-optimal allocation depends on the hazard rate of the distribution
of values.



\paragraph{Consumer-Optimal Information Structures}

We now examine the consumer-optimal information structure where the choice
variable is the quantile value function $W:$ 
\begin{equation}
\max_{W\prec V}U(W,X),  \label{cionsa}
\end{equation}%
where as mentioned we assume that $X(0)=0.$ Appealing to the symmetry
argument in \eqref{sym} we obtain that analytically this is the same problem
as the mechanism design problem of maximizing the revenue 
\begin{equation*}
\max_{W\prec _{w}V}R(X,W)
\end{equation*}
studied in Section \ref{sec:mec}, except for the fact that that majorization
here is strict while earlier we had a weak inequality. In this case, this
difference can lead to different predictions. But of course, if were to
insist on an efficient allocation given the information structure then we
turn the weak majorization constraint into a strict majorization constraint
anyhow.

For concreteness, we consider our leading example of a second price auction
for a single good. Then the unique bidder-optimal symmetric information
structure is given by: 
\begin{equation*}
W(t)=%
\begin{cases}
\frac{\int_{0}^{{t^{\ast }}}V(s)ds}{t^{\ast }}, & \text{if }t<t^{\ast }; \\ 
V(t), & \text{if }t\geq t^{\ast };%
\end{cases}%
\end{equation*}%
where the cutoff $t^{\ast }=\frac{3}{4}\in \lbrack 0,1)$ is independent of
the distribution $F_{v}$. Here we obtained the same threshold to determine
the concavification as in Section \ref{examp}. This is not a coincidence as
we are in fact in both cases concavifying exactly the same function (here it
was useful that $Q=V$ so $r_{Q}$ $=r_{V}$). However, we do not obtain the
same solution as in Section \ref{examp} because the weak and strong
majorization constraint make a difference regarding the nature of the
solution in each instance. When optimizing over information structures, it
is not possible to discard buyers, and the solution pools low values. When
optimizing over allocations, it is possible to discard goods, so those
quantiles that are pooled in the consumer-optimal information structure are
discarded when analyzing the seller-optimal allocation. Hence, in this case
we get some pooling of values while in Section \ref{examp} we obtained that
some qualities are discarded, which would have otherwise being pooled.

\section{Robust Predictions and Other Comparative Statics\label{sec:rob}}

This section discusses robust predictions and comparative statics. It
interprets the preceding results as characterizing the set of feasible
welfare outcomes under all possible information structures. The section
derives the welfare frontier in efficient auctions, showing that the optimal
information structure alternates between upper and lower censoring,
depending on the weight placed on revenue versus consumer surplus. Finally,
it examines comparative statics in the number of bidders, showing that the
optimal pooling threshold rises monotonically with $N$ and that the number
of effectively competing bidders remains approximately constant at about
two, even as $N$ grows large.

\subsection{Robust Predictions}

We have so far considered the problem of mechanism design and information
design with a various different objectives. The mechanism design problem
typically has a very natural literal interpretation but the information
design problem frequently does not have a straightforward interpretation. \
We can think of stories where the information design can have a literal
interpretation: perhaps there is a platform that offers information to
sellers in order to maximize the surplus of consumers on their platform
(see\ \cite{bebm15} and others).

However, we became interested in information design via a different, fully
metaphorical, interpretation of the information design problem. \ The first
generation of information economics made highly parametric assumptions about
information. \ These assumptions translated into extreme structural
assumptions in empirical work that are hard to test. \ But we asked: what
are the set of possible outcomes that can arise in a fixed game: see \cite%
{bemo13a}, \cite{bemo16}, \cite{bemo19}. \ For applications, see \cite%
{bebm15}, \cite{behm15}, \cite{bebm17}, \cite{bebm25}. \ For applications
with twists, see \cite{bebm21b} and \cite{behm20}.

\subsection{Robust Predictions in Efficient Auctions}

We begin by considering the environment studied in Section \ref{auctions},
but we now consider all possible linear combinations of revenue and consumer
surplus (possibly with negative weight). Hence we solve: 
\begin{equation}
\max_{W\prec V}\left\{ \ m\left( (1-|\lambda |)R(W,Q)+\lambda U(W,Q)\right)
\right\}  \label{eq:genw}
\end{equation}%
for all $\lambda \in \lbrack -1,1],\ m\in \{-1,1\}$ and $Q(t)=t^{N-1}.$ We
note that $Q(0)=0$ so we do not need to worry about the discontinuity at %
\eqref{cdxz}. If $\lambda \in \lbrack 0,1],\ m=1$ then the objective
function places positive weight on both consumer surplus and revenue; if $%
\lambda \in \lbrack 0,1],\ m=-1$ then the objective function places negative
weight on both consumer surplus and revenue; if $\lambda \in \lbrack -1,0],\
m=1$ then the objective function places positive weight on revenue but
negative weight on consumer surplus; if $\lambda \in \lbrack -1,0],\ m=-1$
then the objective function places positive weight on consumer surplus but
negative weight on revenue.

Using the symmetry observed earlier in \eqref{sym} we have that: 
\begin{equation*}
U(W,Q)=R(Q,W).
\end{equation*}%
We can thus write the social welfare function as follows: 
\begin{equation*}
m\left( (1-|\lambda |)R(W,Q)+\lambda U(W,Q)\right) =m((1-|\lambda |){e}%
_{Q}(0)+\lambda r_{Q}(0))W(0)+\int_{0}^{1}m((1-|\lambda |){e}_{Q}(t)+\lambda
r_{Q}(t))dW(t).
\end{equation*}%
To make the notation more compact, we define the weighted social welfare, or
simply surplus function as: 
\begin{equation*}
S_{\lambda ,m,Q}(t)=m((1-|\lambda |){e}_{Q}(t)+\lambda r_{Q}(t))
\end{equation*}%
The analysis can be done the same way as we did in Section \ref{sec:infode},
but replacing the excess quality function ${e}_{Q}(t)$ in \eqref{dczc} with
the surplus function ${S}_{\lambda ,m,Q}(t)$. It turns out that $S_{\lambda
,m,Q}(t)$ always has at most one inflection point in $[0,1]$.

There are two types of optimal information structures that can track the
whole frontier of the welfare set: \emph{upper and lower censorship}. As we
construct the concavification of $S_{\lambda ,m,Q}(t)$, we always find an
interval of full disclosure and an interval of pooling, thus a binary and
monotone partition. The interval of pooling can be at the top of the
distribution (as when we maximized revenue in Section \ref{examp:inf-rev})
or it can be at the bottom (as when we maximized consumer surplus in Section %
\ref{examp:inf-cons}). We say the optimal information structure is upper
censorship if there exists a threshold quantile $t_{c}$ such that: 
\begin{equation*}
W(t)=%
\begin{cases}
V(t), & \text{if }t\leq t_{c}; \\ 
\frac{\int_{t_{c}}^{1}V(t)}{1-t_{c}}, & \text{if }t\geq t_{c}.%
\end{cases}%
\end{equation*}%
That is, quantiles $t<t_{c}$ learn their values while quantiles above $t_{c}$
only learn that their value is above $V(t_{c}).$ The solution we obtain in
Section \ref{auctions} was upper censorship. We say the optimal information
structure is lower censorship if there exists $t_{c}$ such that: 
\begin{equation*}
W(t)=%
\begin{cases}
\frac{\int_{0}^{t_{c}}V(t)}{t_{c}}, & \text{if }t\leq t_{c}; \\ 
V(t), & \text{if }t\geq t_{c}.%
\end{cases}%
\end{equation*}%
That is, quantiles $t>t_{c}$ learn their values while quantiles below $t_{c}$
only learn that their value is below $V(t_{c}).$ Complete disclosure and no
disclosure are particular instances of upper and lower censorship

\begin{theorem}[Optimal Information Structure for Weighted Social Welfare]
\label{dcxs3}\quad \linebreak The optimal information structure generates
welfare: 
\begin{equation}
\max_{W\prec V}S_{\lambda ,m,Q}=\overline{S}_{\lambda
,m,Q}(0)V(0)+\int_{0}^{1}\overline{S}_{\lambda ,m,Q}(t)dV(t).  \label{cdxa}
\end{equation}%
If $m=1$ and $\lambda \in \lbrack -1,1/2],$ then the optimal information
structure is upper censorship; otherwise, the optimal information structure
is lower censorship.
\end{theorem}

\begin{figure}[h!]
	\centering
	\includegraphics[width=3.1379in,height=2.9761in,keepaspectratio]{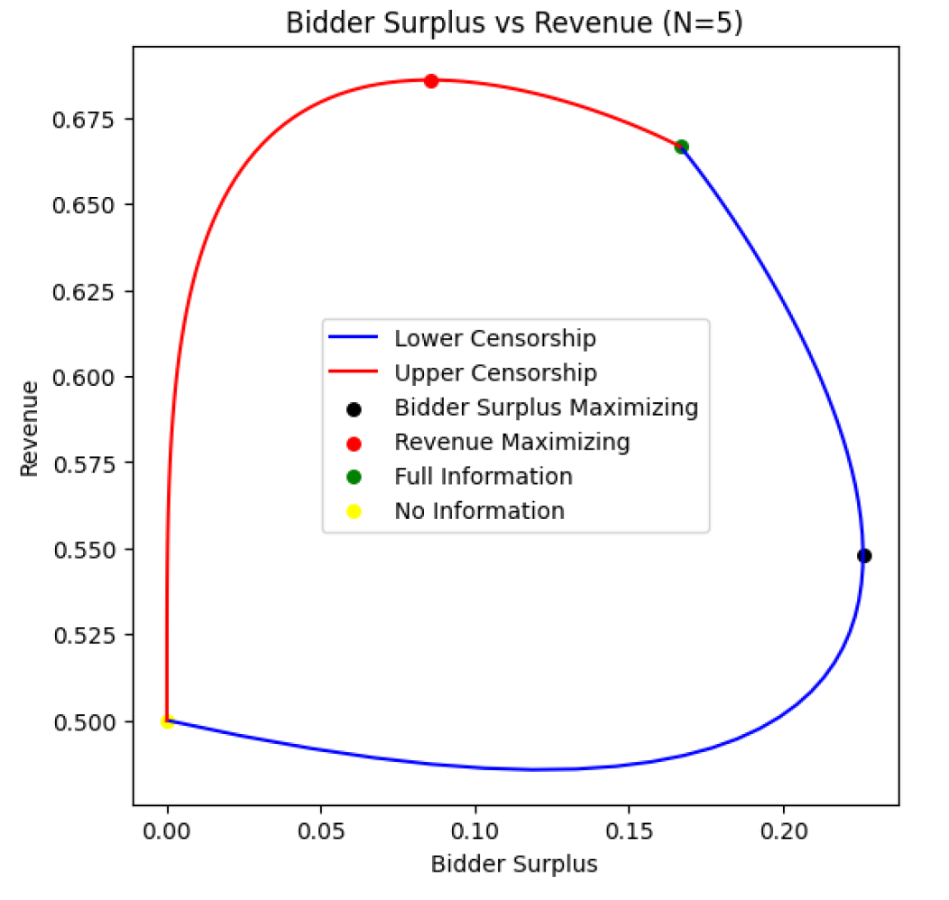}
	\caption{Revenue and Consumer Surplus Frontier across all social welfare weights $\lambda,m$.}
	\label{fig3}
\end{figure}

We now show the boundary of the feasible welfare pairs in Figure \ref{fig3},
where the different colors represent the different solutions. The figure
also shows when full disclosure and when no disclosure is optimal. When $m=1$
and $\lambda =1/2$, then the objective function is to maximize total surplus
so complete disclosure is optimal; $m=-1$ and $\lambda =1/2$, then the
objective function is to minimize total surplus so no disclosure is optimal.
We can see the boundary is non-differentiable at these points, which means
the full disclosure and no disclosure is optimal also for other values close
to $\lambda =1/2.$

\subsection{Allocation and Information Design with Misaligned Objectives}

So far we have either maximized over the allocation or the information while
keeping the other fixed. The only exception was Section \ref{sec:combine} in
which we analyzed the problem when both allocation and information were
chosen with a common objective. Due to the exchange symmetry argument in %
\eqref{sym}, it is easy verify that the same solution would go through if
both were chosen to maximize consumer surplus. However, we may be interested
in situations in which both allocation and information are endogenous to the
game, but they are chosen by different agents with different objectives. An
important class are adversarial or robust setting where the seller chooses
the mechanism and adversarial nature chooses the information structure. We
address this problem in our working paper \cite{behm22a}, \nocite{behm24c}%
(2024). The solution differs significantly from the analysis carried out so
far. In particular, we will not be able to use the results about extreme
points developed by \cite{klms21}, which we presented in Section \ref{sec:om}%
. We now briefly discuss our work in this direction to illustrate some of
novel interpretations and associated results.

To begin, consider the situation in which the information structure is
chosen to maximize consumer surplus and then, after knowing what the
information structure is, the seller chooses the allocation to maximize
revenue. This has a natural literal interpretation where a third party
platform can modify the information provided about consumers, but then
sellers can respond optimally to the demand they face. The problem can be
written as follows: 
\begin{equation}
\max_{\substack{ W\prec V  \\ X\prec _{w}Q}}U(W,X),\text{ }  \label{cop}
\end{equation}%
subject to 
\begin{equation*}
\text{ }X\in \underset{X\prec _{w}Q}{\arg \max }\ R(W,X).
\end{equation*}%
If $Q(t)=t^{N-1}$ then the problem can be interpreted as finding the
consumer-optimal information structure when the seller uses the
revenue-maximizing mechanism (knowing the information structure). If $Q(t)=1$
for all $t\in \lbrack 0,1]$ then we recover the problem studied by \cite%
{rosz17}.

Now, we could also consider the revenue-minimizing information structure.
The problem can be written as follows: 
\begin{equation}
\min_{\substack{ W\prec V  \\ X\prec _{w}Q}}R(W,X)\text{ subject to }X\in 
\underset{X\prec _{w}Q}{\arg \max }\ R(W,X).  \label{cop2}
\end{equation}%
Here is is unclear whether there is a literal interpretation, but one can
interpret this problem as giving bounds on the revenue that a seller can
generate regardless of the information structure.

Furthermore, one can expect to find a saddle point of \eqref{cop2}, and then
the mechanism would have the property of providing the optimal revenue
guarantee across all information structure. Early examples of this approach
are \cite{bebm16b}; \cite{bebm19} while a more general approach has been
recently developed by \cite{brdu21}, \cite{brdu24} and \cite{brdu25b} and
further papers is surveyed in a chapter of this volume, i.e., \cite{brdu25}.
Consistent with this literature, we show a saddle point exists and provide
the corresponding optimal revenue-guarantee mechanism. In contrast to this
early literature we are able to use more basic intuitions of majorization to
construct the saddle point, hence, bringing the tools of majorization to
finding the optimal robust mechanism.

\subsection{Impact of the Number of Bidders in Auctions}

We now analyze how the optimal information structure in an auction changes
with the number of bidders. That is, we revisit the analysis in Section \ref%
{examp:inf-rev} but we see how the solution changes with $N$. A striking
feature of the analysis is that the optimal information structure can be
characterized solely in terms of the number of bidders. In contrast to the
general case of an exogenous inventory, in this case, the exogenous
inventory distribution has a particularly simple form, summarized by a
single parameter, which allows us to derive results with clear economic
interpretations. By contrast, in general, either in the information design
or mechanism design problem, the optimal ironing procedure depends
intricately on the distribution of values or qualities. This makes it harder
to describe the relevant distribution qualitatively.

The optimal information structure is upper censorship, as in %
\eqref{eq:opt-info}. The quantile threshold $t^{\ast }(N)$ that determines
the optimal information structure is equal to zero when there are two
bidders, increases monotonically with $N$, and converges to one as the
number of bidders tends to infinity. For $N\geq 3$, it is implicitly defined
as the solution to the equation 
\begin{equation*}
{e}_{Q}^{\prime }(t)(1-t)\;=\;e_{Q}(1)-e_{Q}(t),
\end{equation*}%
which is a polynomial of degree $N$ in $q$. Numerical solutions for selected
values of $N$ are reported below:%
\begin{equation*}
\begin{array}{ccccccc}
N & 2 & 3 & 4 & 5 & 10 & 100 \\ 
t^{\ast }\left( N\right) & 0 & 0.25 & 0.46 & 0.58 & 0.81 & 0.98%
\end{array}%
\end{equation*}%
It is possible to verify that the expected number of bidders whose values
are pooled at the top is approximately constant across $N$. The expected
number of bidders whose value is at the top of the distribution is: 
\begin{equation*}
N(1-t^{\ast }(N)).
\end{equation*}%
One can then show that: 
\begin{equation*}
N(1-t^{\ast }(N))\in (1.79,2.25)\approx 2\quad \text{for all }N.
\end{equation*}%
Hence, the optimal information policy consistently maintains competition
among about two bidders at the top of the distribution.

\section{Conclusion\label{sec:con}}

This paper has presented a unified framework for analyzing mechanism design
and information design using the language of majorization and quantile
functions. By working in quantile space, we have shown that both problems
reduce to maximizing linear functionals subject to majorization
constraints---with mechanism design involving weak majorization (allowing
discarding of goods) and information design involving strict majorization
(requiring that all types be informed).

Several key insights emerge from our analysis. First, there is a deep
symmetry between the mechanism design and information design problems. Both
involve concavifying an appropriately defined function, and both lead to
pooling or ironing in optimal solutions. However, the nature of the
majorization constraints differs, leading to distinct economic
interpretations: mechanism design excludes low types through reserve prices,
while information design pools their information.

Second, when the designer can optimize over both information and allocation
simultaneously, pooling becomes ubiquitous. Unlike the separate optimization
problems---where full separation or full disclosure can be optimal under
appropriate conditions---joint optimization always involves some pooling.
This reflects the bilinear structure of the objective and the interaction
between the two majorization constraints. Intuitively, even small amounts of
information pooling generate second-order gains in extracting surplus from
inframarginal types, while creating only third-order allocative distortions.

Third, our framework provides a natural approach to robust predictions and
welfare analysis. By characterizing the set of feasible outcomes under all
possible information structures, we can identify predictions that hold
regardless of informational assumptions. In the auction context, we have
shown that the welfare frontier can be completely characterized by upper and
lower censoring policies, with complete disclosure being optimal when total
surplus is maximized.

Fourth, the comparative statics in auctions reveal a striking regularity: as
the number of bidders grows, the optimal information structure pools a
decreasing fraction of the value distribution, yet maintains competition
among approximately two bidders at the top. This suggests that marginal
competition, rather than total competition, is the relevant force in
determining optimal disclosure.

Our analysis can be extended in several directions. First, while we have
focused on exogenous inventories, the framework naturally accommodates
endogenous production decisions with convex costs, as in the original \cite%
{muro78} model. Second, although we have emphasized symmetric settings for
tractability, the techniques extend to asymmetric environments, as shown by 
\cite{bepe07} for auctions. Third, our approach to consumer surplus
maximization and weighted welfare objectives opens connections to platform
design and regulation, where the designer may not fully control both
information and allocation rules.

More broadly, our framework contributes to the emerging second generation of
information economics. The first generation, beginning with \cite{aker70}, 
\cite{spen73}, and \cite{rost76}, employed highly parametric models of
information. Empirical work incorporating private information therefore
relied on strong structural assumptions. The development of data-intensive
tools and the digital economy has both increased the practical importance of
information design and made fully parametric approaches less tenable. Recent
theoretical work has increasingly shifted from simple parametric models to
allowing for general information structures, both in models of costly
information acquisition (\cite{sims03}) and in Bayesian persuasion and
information design (\cite{kage11}; \cite{bemo19}).

Our contribution has been to show that this generality need not come at the
cost of tractability. By working in quantile space and exploiting the
geometry of majorization, we can characterize optimal mechanisms and
information structures, derive robust predictions, and compute welfare
frontiers---all without making restrictive parametric assumptions about the
underlying distributions. The key is recognizing that majorization provides
the right mathematical structure for capturing feasibility constraints in
both mechanism design and information design problems.

The techniques we have developed---concavification, ironing, and the
characterization of extreme points---have classical antecedents in auction
theory and optimal taxation. What is new is the recognition that these same
tools apply, with appropriate modifications, to information design problems.
Moreover, when mechanism and information design are combined, the bilinear
structure leads to qualitatively different conclusions: pooling becomes
optimal even when the primitives would not suggest it in either problem
separately.

As markets become increasingly data-driven and platforms gain greater
ability to control both pricing and information provision, understanding the
joint optimization of these instruments becomes ever more important. Our
framework provides a foundation for analyzing these problems and for
deriving welfare implications that are robust to specific informational
assumptions. The finding that joint optimization leads systematically to
pooling---and the robustness of this result across different
specifications---suggests that concerns about excessive information
manipulation may be warranted when designers control multiple instruments.

\bigskip

\section{Appendix}

\textbf{Proof of Theorem \ref{dcxs}.} We want to maximize: 
\begin{equation}
\max_{X\prec _{w}Q}\left\{ X(0)r_{W}(0)+\int_{0}^{1}r_{W}(t)dX(t)\right\} .
\label{linf}
\end{equation}%
We cannot apply Proposition 2 in \cite{klms21} because we have a weak
majorization instead of a strong majorization. However, with minor
adjustments, we can use their results to identify the optimal solution. We
first define: 
\begin{equation*}
Q_{\theta }(t)=Q(t)\cdot \mathbbm{1}\{t\geq \theta \}.
\end{equation*}%
That is, $\left\{ Q_{\theta }\right\} _{\theta \in \left[ 0,1\right] }$ is
the set of functions that equal $Q(t)$ at quantiles larger than $\theta $
and is zero otherwise. Since \eqref{linf} is a linear maximization over a
convex domain, it attains its optimum at one of the extreme points of the
domain. Following Corollary 2 in \cite{klms21}, the set of extreme points of
the set: 
\begin{equation*}
\{X:X\prec _{w}Q\},
\end{equation*}%
is the union of all the extreme points of the sets of the form: 
\begin{equation*}
\{X:X\prec Q_{\theta }\},
\end{equation*}%
for some $\theta \in \lbrack 0,1]$. That is, we change the weak majorization
for a strong majorization and we change $Q$ for $Q_{\theta }$ and then take
the union over all $\theta .$ Hence, we have that: 
\begin{equation*}
\max_{X\prec _{w}Q}X(0)r_{W}(0)+\int_{0}^{1}r_{W}(t)dX(t)=\max_{\theta \in
\lbrack 0,1]}\left\{ \max_{X\prec Q_{\theta
}}X(0)r_{W}(0)+\int_{0}^{1}r_{W}(t)dX(t)\right\} .
\end{equation*}%
We can now apply Proposition 2 in \cite{klms21} to the innermost
maximization. We obtain: 
\begin{align}
\max_{X\prec Q_{\theta }}X(0)r_{W}(0)+\int_{0}^{1}r_{W}(t)dX(t)=& Q_{\theta
}(0)\overline{r}_{W}(0)+\int_{0}^{1}\overline{r}_{W}(t)dQ_{\theta }(t) 
\notag \\
=& Q(\theta )\overline{r}_{W}(\theta )+\int_{\theta }^{1}\overline{r}%
_{W}(t)dQ(t).  \label{eq:dx2}
\end{align}%
The first equality corresponds to inequality (16) in \cite{klms21} (which in
the same proof is shown to be satisfied with equality); the second equality
is just writing the integral explicitly. By construction, $\overline{r}_{W}$
is concave, so \eqref{eq:dx2} is maximized at $\theta =t_{m}$: 
\begin{equation*}
t_{m}\in \underset{\theta \in \lbrack 0,1]}{\arg \max }\ Q(\theta
)r_{W}(\theta )+\int_{\theta }^{1}\overline{r}_{W}(t)dQ(t).
\end{equation*}%
where we also used that $r_{W}(t_{m})=\overline{r}_{W}(t_{m})$. $%
\blacksquare $

\medskip

\textbf{Proof of Theorem \ref{dcxs2}.} We seek to maximize: 
\begin{equation*}
\max_{W\prec V}\left\{ e_{X}(0)W(0)+\int_{0}^{1}e_{X}(t)dW(t)\right\} .
\end{equation*}%
Now, we cannot directly apply Proposition 2 in \cite{klms21} because $%
e_{X}(t)$ is discontinuous at $t=0.$ However, we have that: 
\begin{equation*}
\lim_{t\downarrow 0}e_{X}(t)<e_{X}(0).
\end{equation*}%
Hence, $e_{X}(t)$ is upper-hemicontinuous; thus, the proof in \cite{klms21}
proceeds in the same way. More precisely, we have that: 
\begin{equation*}
\lim_{t\downarrow 0}e_{X}(t)<e_{X}(0).
\end{equation*}%
Hence, there exists $t_{1}>0$ such that $\overline{e}_{X}(t)>e_{X}(t)$ for
all $t\in \lbrack 0,t_{1}]$. We can thus apply Proposition 2 in \cite{klms21}
by considering a limit of continuous functions. In particular, there exists
a family of functions $\{e_{\ell }(t)\}_{\ell \in \lbrack 0,1]}$, that are
continuously differentiable such that 
\begin{equation*}
e_{\ell }(t)%
\begin{cases}
\in \lbrack e_{X}(t),\overline{e}_{X}(t)], & \text{ for all $t\in \lbrack
0,t_{1}];$} \\ 
=e_{X}(t), & \text{ for all $t\in \lbrack t_{1},1];$}%
\end{cases}%
\end{equation*}%
such that in the limit as $\ell \rightarrow 1$, the functions converge to $%
e_{X}$: 
\begin{equation*}
\lim_{l\rightarrow 1}e_{\ell }(t)=e_{X}(t).
\end{equation*}%
By construction, we have that for all $\ell \in \lbrack 0,1]:$ 
\begin{equation*}
\overline{e}_{\ell }(t)=\overline{e}_{X}(t).
\end{equation*}%
Since the solution is defined solely in terms of the support of the
concavification, we have that the optimal solution is the same for all $\ell
\in \lbrack 0,1]$ and is the same as in Theorem 2. Hence, this is also the
solution for $e_{X}$.$\ \blacksquare $

\newpage

\bibliographystyle{econometrica}
\bibliography{general}

\end{document}